\documentclass[3p]{elsarticle}

\usepackage{lineno,hyperref}

\usepackage{graphicx}
\usepackage{amsmath}
\usepackage{amssymb}
\usepackage{enumerate} 
\usepackage{algorithm}
\usepackage{algcompatible}
\floatname{algorithm}{Pseudocode}

\usepackage{caption}
\usepackage{tabularx}
\usepackage{float}

\newtheorem{theorem}{Theorem}

\modulolinenumbers[5]

\journal{Journal of \LaTeX\ Templates}

\biboptions{authoryear}

\begin{document}

\begin{frontmatter}

\title{A novel Bayesian approach for variable selection in linear regression models}

\author{Konstantin Posch\corref{fn1}}
\ead{Konstantin.Posch@aau.at}

\author{Maximilian Arbeiter\corref{}}
\ead{Maximilian.Arbeiter@aau.at}

\author{Juergen Pilz\corref{}}
\ead{Juergen.Pilz@aau.at}

\cortext[fn1]{Corresponding author}
\address{Universitaet Klagenfurt, Austria}

\begin{abstract}
We propose a novel Bayesian approach to the problem of variable selection in multiple linear regression models. In particular, we present a hierarchical setting which allows for direct specification of a-priori beliefs about the number of nonzero regression coefficients as well as a specification of beliefs that given coefficients are nonzero. To guarantee numerical stability, we adopt a $g$-prior with an additional ridge parameter for the unknown regression coefficients. In order to simulate from the joint posterior distribution an intelligent random walk Metropolis-Hastings algorithm which is able to switch between different models is proposed. Testing our algorithm on real and simulated data illustrates that it performs at least on par and often even better than other well-established methods. Finally, we prove that under some nominal assumptions, the presented approach is consistent in terms of model selection.
\end{abstract}


\end{frontmatter}

\linenumbers

\section{Introduction}
\label{intro}

Nowadays, where computing becomes less expensive, huge amounts of data are routinely collected by companies and organizations of all kinds. The scale of the data being collected increases greatly and thus also the number of measured features (variables/predictors/covariates). Fitting statistical models, which include a large number of features, often results in over-fitting since the resulting model complexity cannot be supported by the available sample size (\cite{Buehlmann}). Moreover, it becomes hard to detect the most predictive features and their importance with respect to the model. Thus, variable selection (i.e. the extraction of the relevant features for a given task) plays an ever more important role in many fileds including genetics (\cite{GeneSelection}), astronomy (\cite{Zheng2007}), and economics (\cite{Foster2004}).

In this paper we are interested in variable selection in the multiple linear regression model
\begin{equation}
\mathbf{y} = \mathbf{X}\boldsymbol{\beta}+\boldsymbol{\varepsilon}
\end{equation}
where $\mathbf{y}$ is a $n$-dimensional response vector, $\mathbf{X}=(\mathbf{x}_1,...,\mathbf{x}_{p})\in\mathbb{R}^{n\times p}$ denotes the so-called design matrix which holds the $p$ potential predictors, $\boldsymbol{\beta}=(\beta_1,...,\beta_p)^T$ is the vector of unknown regression coefficients, and $\boldsymbol{\varepsilon}\sim\mathcal{N}(\mathbf{0},\sigma^2\mathbf{\text{I}}_n)$ denotes the noise vector with independent and identically distributed components. To avoid the need for an intercept $\beta_0$ it is assumed that the response $\mathbf{y}$ is zero centered. Moreover, to allow for an easy evaluation of the influence of single predictors $x_1,...,x_p$ on the model based on the magnitude of the regression coefficients the predictors are treated as zero centered $\sum_{i=1}^nx_{ji}=0$ and standardized with $||\mathbf{x}_j||²=n-1$ for $1\leq j \leq p$.

In the linear model $(1)$ one assumes that only some regression coefficients are different from zero. Thus, the problem of variable selection reduces to the identification of the nonzero regression coefficients (\cite{Alhamzawi2018}). Especially shrinkage approaches such as the lasso (\cite{Tibshirani1996}), the adaptive lasso (\cite{Zou2006}), the elastic net (\cite{Hastie2005}) and their Bayesian analogues (\cite{Park2008,Alhamzawi2012,Leng2014,Huang2015}) which simultaneously perform variable selection and coefficient estimation have been shown to be effective and are often the methods of choice in linear regression. These methods estimate $\boldsymbol\beta$ as minimizer of the objective function $\mathcal{L}(\boldsymbol{\beta})+\mathcal{P}(\boldsymbol\beta,\boldsymbol\lambda)$, where $\mathcal{L}$ denotes the quadratic loss function (negative log-likelihood) $\sum_{i=1}^{n}(y_i-\textbf{x}_i^T\boldsymbol\beta)^2$ and $\mathcal{P}$ denotes a method specific penalty function that encourages a sparse solution. The parameter vector $\boldsymbol\lambda$ controls the penalization strength and thus the level of sparsity. Commonly, a good choice for $\boldsymbol\lambda$ is determined via cross-validation in practical applications.  Among above mentioned approaches the most popular one is the least absolute shrinkage and selection operator (lasso), which was proposed by \cite{Tibshirani1996}. In the lasso regression the penalty function is defined as $\mathcal{P}(\boldsymbol\beta,\lambda):=\lambda ||\boldsymbol\beta||_1=\lambda\sum_{i=1}^p|\beta_i|$. Thus, the regression coefficients are continuously shrunken to zero and for sufficiently large $\lambda$ some coefficients take exactly the value zero. The lasso can be viewed as a convex and therefore more efficiently solvable reformulation of the best subset selection approach, in which the penalty function is given by $\mathcal{P}(\boldsymbol\beta,\lambda):=\lambda ||\boldsymbol\beta||_0=\lambda\#(i|\beta_i\neq 0)$. Moreover, the lasso estimate for $\boldsymbol\beta$ can be interpreted as a Bayesian posterior mode estimate when independent Laplace priors all with zero mean and the same scale parameter $\lambda>0$ are assigned to the regression coefficients:
$$p(\boldsymbol\beta|\sigma^2)=\prod\limits_{i = 1}^{p}\frac{\lambda}{2\sigma}e^{-\lambda\frac{|\beta_i|}{\sigma}}$$
In contrast to the frequentist approach the Bayesian one (\cite{Park2008}) provides credible intervals for the model parameters which can be used for variable selection. A further generalization of the classical lasso is the adaptive lasso (\cite{Zou2006}), which allows for different penalization factors of the regression coefficients:
$$\mathcal{P}(\boldsymbol\beta,\boldsymbol\lambda):=\sum\limits_{i=1}^{p}\lambda_i|\beta_i|$$
For instance, the penalization vector $\boldsymbol\lambda = (\lambda_1,...,\lambda_p)^T$ can be chosen based on the ordinary least squares estimator $\lambda_j^{-1} = |\widehat{\boldsymbol\beta}_j(\text{ols})|^\gamma=|((\mathbf{X}^T\mathbf{X})^{-1}\mathbf{X}^T\mathbf{y})_j|^{\gamma}$ with $\gamma > 0$, or in more complicated settings (collinear predictors, or $n<p$) similarly based on the minimizer of the ridge-regularized problem given by:
$$\underset{\boldsymbol\beta}{\operatorname{min}} \sum\limits_{i=1}^{n}(y_i-\textbf{x}_i^T\boldsymbol\beta)^2+\lambda||\boldsymbol\beta||_2^2$$
The adaptive lasso was proposed to address potential weaknesses of the classical lasso. It has been shown (\cite{Zou2006}) that there are situations in which the classical lasso either tends to select inactive predictors, or over-shrinks the regression coefficients of correct predictores. In contrast, the adaptive lasso satisfies the so-called oracle properties introduced by  \cite{Fan2001}. A procedure satisfies these properties, if it asymptotically identifies the right subset model and has optimal estimation rate. For more details please refer to \cite{Fan2001}. Recently, a Bayesain version of the adaptive lasso was proposed by \cite{Alhamzawi2012} and also by  \cite{Leng2014}. The Bayesian adaptive lasso generalizes the classical (i.e. non adaptive) Bayesian lasso by allowing different scale parameters in the Laplace priors of the regression coefficients:
$$p(\boldsymbol\beta|\sigma^2)=\prod\limits_{i = 1}^{p}\frac{\lambda_i}{2\sigma}e^{-\lambda_i\frac{|\beta_i|}{\sigma}}$$
The scale parameters $\lambda_i$ are then either choosen via marginal maximal likelihood in an emprical Bayesian setting or by assigning appropriate hyperpriors in a hierarchical Bayes approach. Another generalization of the classical lasso is the elastic net proposed by \cite{Hastie2005}. Here the penalty function is given by:
$$\mathcal{P}(\boldsymbol\beta,\boldsymbol\lambda):=\lambda_1 ||\boldsymbol\beta||_1 + \lambda_2 ||\boldsymbol\beta||_2^2$$
The elastic net encourages a grouping of strongly correlated predictors, such that they tend to be in or out the model together. Morover, it works better than the classical lasso in the case where $p\gg n$. There are also some Bayesian versions of the elastic net proposed in literature including the one proposed by \cite{Huang2015}. In particular, we used their R-package \textit{EBglmnet} to validate our novel method. 

Besides above decribed shrinkage approaches, Bayesian methods that are based on the introduction of a random indicator vector $\boldsymbol\gamma=(\gamma_1,...,\gamma_p)^T\in\{0,1\}^p$ gain increasing popularity (\cite{Liang2008,Yongtao2011,Fisher2014,Wang2015}). If $\gamma_i$  equals zero the regression coefficient $\beta_i$ of the $i$-th predictor is assumed to equal zero. This is equivalent to the assumption that the $i$-th predictor does not explain the target $y$. A common choice is to assign independent Bernoulli priors to the indicator variables:
\begin{equation}
p(\boldsymbol\gamma)=\prod\limits_{i=1}^p\gamma_i\pi_i+(1-\gamma_i)(1-\pi_i)
\end{equation}
For instance, \cite{Yongtao2011} used prior $(2)$ with the simplification $\pi =\pi_1=...=\pi_p$ and assigned an uniform Prior $U(a,b)$ to $\operatorname{ln}(\pi)$. In the work of \cite{Fisher2014} the flat prior $\boldsymbol\gamma\propto 1$ was used which is equivalent to seting $\pi_i=0.5$ in $(2)$ for $i=1,...,p$. For given $\boldsymbol\gamma$ the vector of non-zero coefficients denoted by $\boldsymbol\beta_{\boldsymbol\gamma}$ is usally assigned a conventional $g$-prior $\boldsymbol\beta_{\boldsymbol\gamma}|g,\sigma^2,\mathbf{X}_{\mathcal{A}}\sim\mathcal{N}(\mathbf{0},g\sigma^2(\mathbf{X}_{\boldsymbol\gamma}^T\mathbf{X}_{\boldsymbol\gamma})^{-1})$ (introduced by \cite{Zellner1986}) where $\mathbf{X}_{\boldsymbol\gamma}$ denotes the design matrix with all columns deleted that correspond to indicator variables being equal to zero. The reason for the common choice of this prior is that it often leads to a computationally tractable Bayes factor.

Inspired by above described approaches, we propose a setting which is not based on a random indicator vector $\boldsymbol\gamma$, but on a random set $\mathcal{A}\subseteq\{1,...,p\}$ that holds the indices of the active predictors, i.e. the predictors with coefficients different from zero. We assign a prior to $\mathcal{A}$ which depends on the cardinality of the set $|\mathcal{A}|$ as well as on the actual elements of $\mathcal{A}$. This enables an easy formulation of useful a priori knowledge. In section \ref{exp} we show how the results of other methods can be used in order to define such a useful prior in terms of an empirical Bayes approach. To simulate from the joint posterior of our model we propose a novel random walk Metropolis-Hastings algorithm, in contrast to \cite{Yongtao2011,Fisher2014,Wang2015} where Gibbs sampling is applied for this task. This is especially useful when it is difficult or even impossible to determine the conditional distributions needed for the Gibbs sampling algorithm in an analytically closed form. In these cases the corresponding integrals have to be approximated and in particular for practitioners it can become a difficult task to choose a proper approximation technique among the various possible ones. For instance, in \cite{Liang2008} the Laplace approximation is used for this task. \cite{Wang2015} assigned on purpose a beta-prime prior with specific parameters to g in order to achive a closed-form expression of the marginal posterior of $\boldsymbol\gamma$. Since we don't need to care about closed-form expressions by using a random walk Metropolis algorithm, our approach is based on the more popular Zellner-Siow prior for $g$ (\cite{Zellner1980}), which is an inverse gamma distribution with shape parameter $\alpha=1/2$ and scale parameter $\beta = n/2$.

However, the $g$-prior depends on the inverse of the empirical covariance matrix of the selected predictors. This matrix is singular if the number of selected covariates is greater than the number of observations $n$ and, further, may be almost rank deficient given that the predictors are highly correlated. To overcome this problem \cite{Wang2015} replaced the classical inverse with the Moore-Penrose generalized inverse and thus ended up with the so-called $gsg$-prior (see \cite{West2003}). In contrast to them, we adopt a $g$-prior with an additional ridge parameter for the unknown regression coefficients to guarantee nonsingularity of the empirical covariance matrix. This modification of the classical $g$-prior was first proposed by  \cite{Gupta2007} and further investigated by \cite{Baragatti2012}.

Finally, in Section \ref{consistency} we state that our approach is consistent in terms of model selection according to the consistency definition given by \cite{FERNANDEZ2001}. The proof of this result is deferred to the appendix. Moreover, in Section \ref{exp}, we evaluate our approach on the basis of real and simulated data and compare the results with the already described shrinkage methods. We show that our approach performs at least on par and often better than the comparative methods.


\section{Methodology}
\label{methodology}

In this section we describe the hierarchical Bayesian approach we propose for the task of variable selection in multiple linear regression models. Moreover, we state that the novel model presented is consistent in terms of model selection. Finally, we propose an intelligent random walk Metropolis-Hastings algorithm to simulate from the joint posterior of the model parameters.

\subsection{Hierarchical representation of the full model}
\label{model}

In order to perform variable selection, a random set $\mathcal{A}\subseteq\{1,...,p\}$ which holds the indices of the active predictors, i.e. the predictors with coefficients different from zero, is used. Moreover, it is assumed that the cardinality $k=|\mathcal{A}|$ is greater than zero. In this paper the focus lies on problems where at least one of the measured features is predictive. Nevertheless, one can easily generalize the proposed approach in such a way that also the null model  (all regression coefficients equal zero) is valid. Thus, we propose the following prior for $\mathcal{A}$:
\begin{equation}
p(\mathcal{A}=\{\alpha_1,...,\alpha_{k}\})\propto (p_{\alpha_1}+...+p_{\alpha_k})\frac{1}{k}\tilde{p}(k)
\end{equation}
where
\begin{enumerate}[(i)]
\item $\{p_{\alpha_1},...,p_{\alpha_k}\}\subseteq\{p_1,...,p_p\}$ with $\sum_{i=1}^{p}p_i=1$ and $p_i\geq 0$ for $i=1,...,p$
\item $\tilde{p}:\{1,...,p\}\rightarrow\mathbb{R}_{0}^{+}$.
\end{enumerate}
The mapping $\tilde{p}$ can be arbitrarily chosen and should represent the a-priori belief in the model size, i.e. the number of parameters different from zero. For instance one could define $\tilde{p}$ based on the probability mass function of a zero truncated binomial distribution with size parameter equal to the number of predictors $p$ and the second parameter chosen in such a way that the mean of the distribution equals the a-priori most probable model size. On the other hand the parameters $p_1,...,p_p$ are used to represent the a-priori belief in the importance of the predictors $x_1,...,x_p$ to the model. In terms of empirical Bayes these parameters could be chosen based on the regression coefficients of a ridge regularized model (compare to the adaptive lasso) or based on the correlations (i.e. the linear dependencies) between the target variable and the predictors.  Please note that in the case of equal a-priori importance of the predictors $\frac{1}{p}=p_1=...=p_p$, the prior $(3)$ reduces to
$$p(\mathcal{A}=\{\alpha_1,...,\alpha_{k}\})\propto\tilde{p}(k).$$
Thus, in this case the prior only reflects the a-priori belief regarding $k$, which shows the necessity of the factor $\frac{1}{k}$ in $(3)$.

For the variance $\sigma^2$ of the error terms $\varepsilon_1,...,\varepsilon_n$ an inverse gamma prior is chosen:
\begin{equation}
p(\sigma^2)\propto (\sigma^2)^{-(a+1)}\operatorname{exp}\left(-\frac{b}{\sigma^2}\right)
\end{equation}
For $b\rightarrow 0$ and $a\rightarrow 0$ the inverse gamma prior converges to the non-informative and improper Jeffreys prior $p(\sigma^2)\propto \frac{1}{\sigma^2}$ (\cite{Jeffrey1961}) which is invariant under a change of scale. In the experimental studies in Section \ref{exp}, $a$ and $b$ are both set to $0.001$ such that the chosen prior is a proper approximation of the Jeffreys prior.

As already stated in the introduction, for given $\mathcal{A}$ the vector of non-zero coefficients denoted by $\boldsymbol\beta_{\mathcal{A}}$ is commonly assigned a conventional $g$-prior $\boldsymbol\beta_{\mathcal{A}}|g,\sigma^2,\mathbf{X}_{\mathcal{A}}\sim\mathcal{N}(\mathbf{0},g\sigma^2(\mathbf{X}_{\mathcal{A}}^T\mathbf{X}_{\mathcal{A}})^{-1})$ (introduced by \cite{Zellner1986}), where $\mathbf{X}_{\mathcal{A}}$ denotes the submatrix of $\mathbf{X}$ consisting of all columns corresponding to predictors with index in $\mathcal{A}$. Since the $g$-prior depends on the inverse of the empirical covariance matrix of the selected predictors it is singular if the number of selected covariates $k$ is greater than the number of observations $n$ and further may be almost rank deficient provided that the predictors are highly correlated (see \cite{West2003,Gupta2007,Baragatti2012}). To overcome these two problems, based on the ideas of \cite{Gupta2007} and  \cite{Baragatti2012}, we consider a ridge penalized version of the $g$-prior:
\begin{equation}
\boldsymbol\beta_{\mathcal{A}}|g,\sigma^2,\mathbf{X}_{\mathcal{A}}\sim\mathcal{N}(\mathbf{0},(g^{-1}\sigma^{-2}\mathbf{X}_{\mathcal{A}}^T\mathbf{X}_{\mathcal{A}}+\lambda \mathbf{\text{I}}_{k})^{-1})
\end{equation}
where
\begin{equation}
\lambda = \begin{cases}
\operatorname{max}(\frac{1}{k},\frac{1}{300}) &\text{if~} n\leq\zeta\\
0 & \text{else}
\end{cases}
\end{equation}
with $\zeta$ denoting an appropriately large constant. It should be explained what appropriately large means. Obviously, for $n>\zeta = p-1$ the first possible problem ($k>n$) cannot appear. The second problem, i.e. an almost rank deficient matrix $\mathbf{X}_{\mathcal{A}}^T\mathbf{X}_{\mathcal{A}}$, also should not appear for $n\gg k$, i.e. for sufficiently large $\zeta$. Even if the predictors are highly correlated, a huge sample size will prevent $\mathbf{X}_{\mathcal{A}}^T\mathbf{X}_{\mathcal{A}}$ from beeing nearly singular and thus maybe computationally singular (compare \cite{Carsey2013}). Although it has to be assumed that there is no predictor that is an exact linear combination of the other predictors. However, this assumption can be easily achieved by excluding such non-informative predictors from the beginning. It should be mentioned that setting $\lambda$ to zero is not necessary at all in practical applications. This is just a theoretical consideration which will ease the proof of model selection consistency in \ref{appendix}. The definition of $\lambda$ for $n\leq \zeta$ is taken from \cite{Baragatti2012}, who suggested to reduce the influence of the ridge parameter when the number of predictors is large up to the threshold $1/300$.

Finally, we assign the popular Zellner-Siow prior (\cite{Zellner1980}), i.e. an inverse gamma distribution with shape parameter $\alpha=1/2$ and scale parameter $\beta = n/2$ to $g$, resulting in the complete hierarchical representation of the model
\begin{equation}
\mathbf{y}|\mathbf{X},\boldsymbol{\beta},\sigma^2\sim\mathcal{N}(\mathbf{X}\boldsymbol{\beta},\sigma^2\mathbf{\text{I}}_n)
\end{equation}
\begin{align}
p(\boldsymbol\beta|\mathcal{A},g,\sigma^2,\mathbf{X}_{\mathcal{A}})\propto\operatorname{exp}\left(-\frac{1}{2}\boldsymbol\beta_{\mathcal{A}}^T \left(g^{-1}\sigma^{-2}\mathbf{X}_{\mathcal{A}}^T\mathbf{X}_{\mathcal{A}}+\lambda \mathbf{\text{I}}_{k}\right) \boldsymbol\beta_{\mathcal{A}}\right)h(\boldsymbol\beta_{\mathcal{A}^C})
\end{align}
\begin{equation}
p(\mathcal{A}=\{\alpha_1,...,\alpha_{k}\})\propto (p_{\alpha_1}+...+p_{\alpha_k})\frac{1}{k}\tilde{p}(k)
\end{equation}
\begin{equation}
g\sim\text{IG}\left(\frac{1}{2},\frac{n}{2}\right)
\end{equation}
\begin{equation}
\sigma^2\sim\text{IG}\left(a,b\right)
\end{equation}
where $\boldsymbol\beta_{\mathcal{A}^C}$ denotes the components of $\boldsymbol\beta$ with indices not in $\mathcal{A}$,
\begin{equation}
h(\boldsymbol\beta_{\mathcal{A}^C}) =
\begin{cases}
 1 &\text{if~}\boldsymbol\beta_{\mathcal{A}^C}=\boldsymbol{0}\\
 0 &\text{else}
\end{cases}
\end{equation}
and $\lambda$ satisfies equation $(6)$.

\subsection{Model consistency}
\label{consistency}

In this section we state that model $(7-12)$ is consistent in terms of model selection. The proof of the result is is given in \ref{appendix}. Therefore, throughout the section it is assumed that the sample $\mathbf{y}$ is generated by model $M_{\mathcal{A}}$ with parameters $\boldsymbol\beta_{\mathcal{A}}$ and $\sigma^2$, i.e.,
\begin{equation}
\mathbf{y}=\mathbf{X}_{\mathcal{A}}\boldsymbol\beta_{\mathcal{A}}+\boldsymbol\varepsilon \text{ ~with } \boldsymbol\varepsilon\sim\mathcal{N}(\mathbf{0},\sigma^2\mathbf{\text{I}}_n).
\end{equation}
In rough words, model selection consistency means that the true model will be selected provided that enough data is available. A sound mathematical definition was given by \cite{FERNANDEZ2001}. According to these authors the posterior probability is said to be consistent if
\begin{equation}
\underset{n\rightarrow \infty}{\operatorname{p~lim~}} p(M_{\mathcal{A}}|\mathbf{y},\mathbf{X})=1 \text{~~~and~~~} \underset{n\rightarrow \infty}{\operatorname{p~lim~}} p(M_{\mathcal{A}^{\prime}}|\mathbf{y},\mathbf{X})=0 \text{~~~for all~}\mathcal{A}^{\prime}\neq \mathcal{A}
\end{equation}
where the probability limit is taken with respect to the true sampling distribution $(13)$. Moreover, some preliminary results which are required in our proof are formulated as Lemma A.1 in their work. In our notation this Lemma reads as:\vspace{0.3cm}

\textit{Under the sampling model (also true model) $M_{\mathcal{A}}$ $(13)$,
\begin{enumerate}
\item If $M_{\mathcal{A}}$ is nested within or equal to model $M_{\mathcal{A}^{\prime}}$,
\begin{equation}
\underset{n\rightarrow \infty}{\operatorname{p~lim~}} \frac{\mathbf{y}^T(\mathbf{\text{I}}_{n}-\mathbf{X}_{\mathcal{A}^{\prime}}\left(\mathbf{X}_{\mathcal{A}^{\prime}}^T\mathbf{X}_{\mathcal{A}^{\prime}}\right)^{-1}\mathbf{X}_{\mathcal{A}^{\prime}}^T)\mathbf{y}}{n} = \sigma^2.
\end{equation}
\item Under the assumption that for any model $M_{\mathcal{A}^{\prime}}$ that does not nest $M_{\mathcal{A}}$,
\begin{equation}
\underset{n\rightarrow \infty}{\operatorname{lim~}} \frac{\beta_{\mathcal{A}}^T\mathbf{X}_{\mathcal{A}}^T(\mathbf{\text{I}}_{n}-\mathbf{X}_{\mathcal{A}^{\prime}}\left(\mathbf{X}_{\mathcal{A}^{\prime}}^T\mathbf{X}_{\mathcal{A}^{\prime}}\right)^{-1}\mathbf{X}_{\mathcal{A}^{\prime}}^T)\mathbf{X}_{\mathcal{A}}\beta_{\mathcal{A}}}{n} = b_{\mathcal{A}^{\prime}}\in(0,\infty),
\end{equation}
one obtains
\begin{equation}
\underset{n\rightarrow \infty}{\operatorname{p~lim~}} \frac{\mathbf{y}^T(\mathbf{\text{I}}_{n}-\mathbf{X}_{\mathcal{A}^{\prime}}\left(\mathbf{X}_{\mathcal{A}^{\prime}}^T\mathbf{X}_{\mathcal{A}^{\prime}}\right)^{-1}\mathbf{X}_{\mathcal{A}^{\prime}}^T)\mathbf{y}}{n} = \sigma^2+ b_{\mathcal{A}^{\prime}}.
\end{equation}
\end{enumerate}}
\vspace{0.3cm}

\begin{theorem}\label{theo1}
Assume that $(16)$ holds. Then the proposed model setup $(7-12)$ is consistent in terms of model selection.
\end{theorem}

\subsection{Metropolis-Hastings algorithm}
\label{metropolis}

As usual in Bayesian regression models, the joint posterior $p(\boldsymbol\beta,\mathcal{A}, g,\sigma^2|\mathbf{y},\mathbf{X})$ is analytically intractable. The most common approach to overcome this problem is to use Markov chain Monte Carlo (MCMC) methods to simulate from this unknown distribution. Such a method generates realizations of a Markov chain which converges in distribution to the (conditional) random vector of interest, i.e. in this paper to $\boldsymbol\beta,\mathcal{A}, g,\sigma^2|\mathbf{y},\mathbf{X}$. Thus, after some time of convergence, the so-called burn-in phase, the generated random numbers can be considered as a (dependent) sample from the posterior distribution. To finally obtain an independent sample only simulated numbers with a given distance $d$ are chosen, a so-called thinning is performed. In this paper a special Metropolis Hastings (MH) algorithm, and thus a special MCMC method, is proposed to simulate from 
$p(\boldsymbol\beta,\mathcal{A}, g,\sigma^2|\mathbf{y},\mathbf{X})$. MH algorithms are iterative and in each iteration a random number is drawn from a proposal distribution $q$ which is then accepted or not based on a given criterion. If the proposal $q$ depends on the respectively last accepted random number one also speaks of a random walk MH algorithm. For further details on MCMC methods please refer to \cite{Fahrmeir2013}. In order to define a MH algorithm for a given problem basically the only thing one has to specify is the proposal distribution $q$ used in the algorithm. Therefore, we define the proposal $q(\boldsymbol\beta_{t+1},\mathcal{A}_{t+1}, g_{t+1},\sigma_{t+1}^2|\mathcal{A}_{t}, g_{t},\sigma_{t}^2)$ as follows:

Inspired by the work of \cite{Hastings1970} the proposal distribution for the variance parameter $\sigma^2$ as well as the proposal distribution of the parameter $g$ are defined to be uniform distributions
\begin{align}
q(\sigma_{t+1}^2|\sigma_{t}^2)&=\text{Unif}\left(\sigma_{t+1}^2;\operatorname{max}\left\{10^{-8},\sigma_{t}^2-\varepsilon_{\sigma}\right\},\sigma_{t}^2+\varepsilon_{\sigma}\right)\\
q(g_{t+1}|g_{t})&=\text{Unif}\left(g_{t+1};\operatorname{max}\left\{10^{-8},g_{t}-\varepsilon_{g}\right\},g_{t}+\varepsilon_{g}\right)
\end{align}
where $\varepsilon_{\sigma}>0$ and $\varepsilon_{g}>0$ denote tuning parameters. Tuning parameters are always chosen in such a way, that the acceptance rate of the algorithm is neither very low nor very high which should guarantee a fast convergence of the algorithm. In order to specify the proposal for the random set $\mathcal{A}$ at first a Bernoulli distributed random variable $c_{h}$ is introduced
\begin{equation}
c_{h}\sim \text{Bernoulli}(p_h)
\end{equation}
where $p_{h}\in [0,1]$ can be considered as a tuning parameter. The event $c_h=1$ means that the model size (number of nonzero regression coefficients) changes, i.e. $k_{t+1}\neq k_t$. Moreover, a random variable $\alpha$ with support on the index set $\{1,...,p\}$ of the regression coefficients is introduced in order to describe the model transition probabilities in case of changing model size. A realization of this random variable corresponds to the index of a predictor which is going to be added to or removed from the model. Transitions between models, which differ by two or more parameters, are not allowed. The probability mass function of $\alpha$ is defined as
\begin{equation}
q(\alpha|\mathcal{A}_t)=\text{I}_{\{1,...,p\}}(\alpha)\begin{cases}
\tilde{p}_{\alpha} &\text{if } k_t>1,\alpha\notin\mathcal{A}_t\\
\frac{\sum\limits_{i\in\mathcal{A}_t}\tilde{p}_i}{\tilde{p}_\alpha\sum\limits_{i\in\mathcal{A}_t}1/\tilde{p}_i}&\text{if } k_t>1,\alpha\in\mathcal{A}_t\\
0&\text{if } k_t=1,\alpha\in\mathcal{A}_t\\
\frac{\tilde{p}_{\alpha}}{\sum\limits_{i\in\{1,...,p\}\setminus\mathcal{A}_t}\tilde{p}_i} &\text{else}
\end{cases}
\end{equation}
where the parameters $\tilde{p}_1,...,\tilde{p}_p$ are greater or equal to zero, sum up to one, and can but must not be identical to the parameters $p_1,...,p_p$ already used in the prior $p(\mathcal{A})$. Again, as the parameters $p_1,...,p_p$ the parameters $\tilde{p}_1,...,\tilde{p}_p$ can be used to represent a-priori beliefs about the importance of the predictors $x_1,...,x_p$ and a good choice of them should lead to a fast convergence of the MH algorithm. Especially, in problems with a high number of predictors it is essential to assign  $\tilde{p}_1,...,\tilde{p}_p$ with informative values, since otherwise it might take a very large number of iterations until a MCMC method (see Section \ref{metropolis}) used to simulate from the joint posterior might converge. According to $(21)$ in the case where $k_t>1$ a predictor $x_{\alpha}$ is added with probability $p_{\alpha}$ which might correspond to the a-priori importance of this predictor. Furthermore, a predictor $x_{\alpha}$ is removed with a probability proportional to $p_{\alpha}^{-1}$ which might be the inverse a-priori importance of $x_{\alpha}$. In the case where $k_t=1$ a predictor $x_{\alpha}$ is added with a probability proportional to $p_{\alpha}$ and it is impossible that the only predictor in the model is removed, this would violate the prior assumption that $k=|\mathcal{A}|>0$ with probability $1$. Using $(20)$ and $(21)$ the proposal distribution $q(\mathcal{A}_{t+1}|\mathcal{A}_{t})$ is finally defined by:
\begin{align}
&q(\mathcal{A}_{t+1}|\mathcal{A}_{t},c_h=1)=q([\mathcal{A}_{t+1}\setminus\mathcal{A}_{t}]\cup[\mathcal{A}_{t}\setminus\mathcal{A}_{t+1}]|\mathcal{A}_{t})\text{I}_{\{1\}}(|k_{t+1}-k_t|)\\
&q(\mathcal{A}_{t+1}|\mathcal{A}_{t},c_h=0)=\begin{cases}
1 &\text{if } \mathcal{A}_{t+1}=\mathcal{A}_t\\
0 &\text{else}
\end{cases}\\
&q(\mathcal{A}_{t+1}|\mathcal{A}_{t})=\sum\limits_{c_h\in\{0,1\}}q(\mathcal{A}_{t+1}, c_h|\mathcal{A}_{t})\\
&\hspace{1.5cm}= \sum\limits_{c_h\in\{0,1\}}q(\mathcal{A}_{t+1}|c_h, \mathcal{A}_{t})q(c_h)
\end{align}
Note that the probability mass function $q$ in equation $(22)$ is given by equation $(21)$, since $|k_{t+1}-k_t|=1$. Besides the easy evaluation of $q(\mathcal{A}_{t+1}|\mathcal{A}_{t})$ for a given input, one can easily simulate from this distribution. This is done by sampling from the distribution of $c_h$ followed by sampling from the conditional distribution of $\mathcal{A}_{t+1}|\mathcal{A}_{t},c_h$ where the condioning takes places with respect to the sample of $c_h$ generated in the previous step. The only thing remaining to define the overall proposal $q(\boldsymbol\beta_{t+1},\mathcal{A}_{t+1}, g_{t+1},\sigma_{t+1}^2|\mathcal{A}_{t}, g_{t},\sigma_{t}^2,\mathbf{X})$ is the conditional proposal $q(\boldsymbol\beta_{t+1}|\mathcal{A}_{t+1}, g_{t+1},\sigma_{t+1}^2,\mathbf{X}_{\mathcal{A}_{t+1}})$. It is well known (see \cite{Fahrmeir2013}) that under the observation model $\mathbf{y}|\boldsymbol\beta,\sigma^2,\mathbf{X}\sim\mathcal{N}(\mathbf{X}\boldsymbol\beta,\sigma^2\mathbf{\text{I}})$ with prior distributions $\boldsymbol\beta|\sigma^2~\sim\mathcal{N}(\mathbf{0},\sigma^2\mathbf{M})$ and $\sigma^2\sim \text{IG}(a,b)$ the conditional posterior of $\boldsymbol\beta$ given $\sigma^2$ is given by
\begin{equation*}
\boldsymbol\beta|\sigma^2,\mathbf{y},\mathbf{X}\sim\mathcal{N}(\tilde{\mathbf{m}},\sigma^2\tilde{\mathbf{M}})
\end{equation*}
with
\begin{equation*}
\tilde{\mathbf{m}} = \tilde{\mathbf{M}}\mathbf{X}^{T}\mathbf{y}
\end{equation*}
\begin{equation*}
\tilde{\mathbf{M}} = \left(\mathbf{X}^{T}\mathbf{X}+\mathbf{M}^{-1}\right)^{-1}.
\end{equation*}
Thus, a natural choice for $q(\boldsymbol\beta_{t+1}|\mathcal{A}_{t+1}, g_{t+1},\sigma_{t+1}^2,\mathbf{X}_{\mathcal{A}_{t+1}})$ is given by
\begin{align*}
q(\boldsymbol\beta_{t+1}|\mathcal{A}_{t+1}, g_{t+1},\sigma_{t+1}^2,\mathbf{X}_{\mathcal{A}_{t+1}})&=q(\boldsymbol\beta_{t+1,\mathcal{A}_{t+1}}|\mathcal{A}_{t+1}, g_{t+1},\sigma_{t+1}^2,\mathbf{X}_{\mathcal{A}_{t+1}})q(\boldsymbol\beta_{t+1,\mathcal{A}_{t+1}^{C}}|\mathcal{A}_{t+1})\\
&\propto\operatorname{exp}\left(-\frac{1}{2}\left(\boldsymbol\beta_{t+1,\mathcal{A}_{t+1}}-\boldsymbol{\mu}_{t+1}\right)^T\mathbf{F}_{t+1}\left(\boldsymbol\beta_{t+1,\mathcal{A}_{t+1}}-\boldsymbol{\mu}_{t+1}\right)\right)h(\boldsymbol\beta_{t+1,\mathcal{A}_{t+1}^{C}})
\end{align*}
where
\begin{equation*}
\boldsymbol{\mu}_{t+1} = \sigma_{t+1}^{-2}\mathbf{F}_{t+1}^{-1}\mathbf{X}_{\mathcal{A}_{t+1}}^{T}\mathbf{y}
\end{equation*}
and 
\begin{equation*}
\mathbf{F}_{t+1} = \sigma_{t+1}^{-2}\mathbf{X}_{\mathcal{A}_{t+1}}^{T}\mathbf{X}_{\mathcal{A}_{t+1}}+\sigma_{t+1}^{-2}g_{t+1}^{-1}\mathbf{X}_{\mathcal{A}_{t+1}}^{T}\mathbf{X}_{\mathcal{A}_{t+1}}+\lambda\mathbf{\text{I}}_{k_{t+1}}.
\end{equation*}
Finally, the overall proposal distribution can be written as
\begin{align}
&q(\boldsymbol\beta_{t+1},\mathcal{A}_{t+1}, g_{t+1},\sigma_{t+1}^2|\mathcal{A}_{t}, g_{t},\sigma_{t}^2,\mathbf{X})=\\
&=q(\boldsymbol\beta_{t+1}|\mathcal{A}_{t+1}, g_{t+1},\sigma_{t+1}^2,\mathbf{X}_{\mathcal{A}_{t+1}})q(\mathcal{A}_{t+1}|\mathcal{A}_{t})q(\sigma_{t+1}^2|\sigma_{t}^2)q(g_{t+1}|g_{t}).
\end{align}
One can easily simulate from the overall proposal by iteratively simulating from the factors in $(27)$ from right to left and conditioning on the values sampled up to the current step of execution.

\section{Experimental studies}
\label{exp}

In this section the (predictive) performance of the proposed hierarchical Bayesian approach is analysed and compared with some other Bayesian and non-Bayesian methods. For this task we have implemented the MH algorithm described in Section \ref{metropolis} by using the interface between the programming languages R (\cite{R}) and C. The comparison includes the lasso (\cite{Tibshirani1996}), the adaptive lasso (alasso) (\cite{Zou2006}), the elastic net (elastic) (\cite{Hastie2005}), the Bayesian lasso (blasso) (\cite{Park2008}), the Bayesian adaptive lasso (balasso) (\cite{Alhamzawi2012,Leng2014}) and the Bayesian elastic net (belastic) (\cite{Huang2015}). For our extensive comparisons three real-world datasets and also simulated datasets from two different artificial models are used. The performance comparison on the real-world datasets is carried out on the basis of a $5$-fold cross-validation. Performing a $5$-fold cross-validation on a given dataset means partitioning the dataset in $5$ equal sized subsets, iteratively selecting each of these subsets exactly once as testing data, whilst using  the remaining subsets for training. Aggregating some measure of accuracy, respectively computed from each of the $5$ testing datasets, results in the overall performance of the models to evaluate. The performance comparison for the artificial models is accomplished by simulating multiple datasets from these models, then performing a train/test split for each simulated dataset, and finally again aggregating the single accuracies achieved. The accuracy measures chosen in this paper are the mean squared error (MSE) and the mean absolute deviation (MAD)
\begin{align}
\text{MSE}&=\frac{1}{n_{te}}\sum\limits_{i=1}^{n_{te}}(y_i-\widehat{y}_i)^2\\
\text{MAD}&=\frac{1}{n_{te}}\sum\limits_{i=1}^{n_{te}}|y_i-\widehat{y}_i|
\end{align}
where $n_{te}$ denotes the cardinality of the testing dataset and the $\widehat{y}_i$ denote the predicted target values, for $i=1,...,n_{te}$. For the aggregation of single accuracies we are using the median since it is more robust to outliers than the mean of the values. The median of MSEs is then denoted by MMSE and the median of MADs by MMAD. For our novel approach it should be stressed that, for obtaining a prediction $\widehat{y}^*$ corresponding to a test sample $\mathbf{x}^*$, an estimate of the expected value of the posterior predictive distribution $\mathbb{E}(y^*|\mathbf x^*,\mathbf{y},\mathbf{X})$ is used. Using Monte Carlo integration this expected value can be estimated as follows:
\begin{align*}
\mathbb{E}(y^*|\mathbf{x}^*,\mathbf{y},\mathbf{X})&=\int y^*p(y^*|\mathbf x^*,\mathbf{y},\mathbf{X})~dy^*\\
&=\int y^*\left\{\int\int p(y^*|\mathbf x^*,\boldsymbol\beta,\sigma^2)p(\boldsymbol\beta,\sigma^2|\mathbf{y},\mathbf{X}) ~d\sigma^2d\boldsymbol\beta\right\}~dy^*\\
&=\int\int\left\{\int y^*p(y^*|\mathbf x^*,\boldsymbol\beta,\sigma^2)~dy^*\right\}p(\boldsymbol\beta,\sigma^2|\mathbf{y},\mathbf{X}) ~d\sigma^2d\boldsymbol\beta\\
&=\int\int \mathbf x^{*T}\boldsymbol\beta~ p(\boldsymbol\beta,\sigma^2|\mathbf{y},\mathbf{X})~d\sigma^2d\boldsymbol\beta\\
&=\int \mathbf x^{*T}\boldsymbol\beta~ p(\boldsymbol\beta|\mathbf{y},\mathbf{X})~d\boldsymbol\beta\\
&\approx \frac{1}{N}\sum\limits_{i=1}^{N}\mathbf x^{*T}\boldsymbol\beta_i
\end{align*}
where $\boldsymbol\beta_i,...,\boldsymbol\beta_N$ is a sample of size $N$ from the marginal posterior $p(\boldsymbol\beta|\mathbf{y},\mathbf{X})$. For all the non-Bayesian comparsion models the predicted target values $\widehat{y}_1,...,\widehat{y}_{te}$ are computed by calculating the inner product of the test samples with the estimate $\widehat{\boldsymbol\beta}$ obtained by minimizing the corresponding model-specific objective function. For the Bayesian comparsion models the way the predictions are computed depend on the output provided by the R-packages used in this paper. In this work the R-function \textit{blasso} included in the R-package \textit{monomvn} (\cite{Rblasso}) is applied in order to perform Bayesian lasso regression. Since the function \textit{blasso} returns a sample from the marginal posterior $p(\boldsymbol\beta|\mathbf{y},\mathbf{X})$, as for the model proposed by us, the predictions are computed by estimating the posterior predictive mean. Moreover, we use the R-Function \textit{brq} included in the package \textit{Brq} (\cite{Rbalasso}) for applying the Bayesian adaptive lasso regression. It should be mentioned that this R-function is not developed for the classical linear regression setting where the conditional mean of the response variable is modelled, rather it is developed for the more general setting where any conditional quantile is modelled. For purposes of the comparision study in this paper the conditional median, i.e. the conditional $50\%$-quantile, is selected. The R-function produces a Bayes estimate of $\boldsymbol\beta$ which is then used for prediction (i.e. inner product computation with the test samples). In order to use the Bayesian elastic net the function \textit{EBglmnet} included in the package \textit{EBglmnet} (\cite{REBglmnet}) is used in this work. This function returns the posterior mean of the nonzero regression coefficients, which is then used for prediction (again via inner product computation with the test samples).

It is also important to mention how (inside the $5$-fold cross-validation for real-world datasets, or for the multiple simulated datasets) the hyperparameters of the trained models are determined. For the frequentist models, which are all implemented in the R-package \textit{glmnet} (\cite{Rglmnet}), the R-function \textit{cv.glmnet} is used, which is also included in this package. The penalization factor $\lambda$ of the lasso regression is chosen based on a $10$-fold cross-validation on a one dimensional grid with the MAD as accuracy measure. Analogously, the penalization factors $\lambda_1,\lambda_2$ of the elastic net are selected based on a $10$-fold cross-validation on a now two dimensional grid, again with MAD as accuracy measure. For the adaptive lasso the chosen penalization factors are the absolute values of the inverse regression coefficients obtained from the respective ridge-penalized problem, the penalization factor of which is also determined via $10$-fold cross-validation. For the Bayesian version of the lasso and the  adaptive lasso the default values of the corresponding functions \textit{blasso} and \textit{brq} are assigned to the hyperparameters of the a-priori distributions. For further details please refer to the help pages of these packages. However, it should be mentioned that for the Bayesian lasso the option of additional model selection based on a Reversible Jump MCMC algorithm is set to true for the experimental studies in this paper. Moreover, for the Bayesian elastic net the hyperparameters in the prior distribution are selected via $5$-fold cross-validation. This is done by using the function \textit{cv.EBglmnet}, which is also available in the package \textit{EBglmnet}. The hyperparameters of our proposed Bayesian model are chosen differently in the following experimental studies to indicate useful possible choices. These specific choices are described later on in this article.

In order to obtain i.i.d. samples from the model-specific posterior distributions the R-functions used in this paper merely save a fraction of all samples generated by the MCMC methods implemented. For the Bayesian elastic net, the decision which samples to store is already taken by the implementation itself and cannot be user-defined. All other Bayesian models go along with implementations which allow for a user specific determination of this fraction. To do so the following procedure is chosen:  In a first step, respectively, the first $10,000$ samples are 
deleted and, in a second step, all samples except of every $10$-th one are deleted. Note that samples generated in the first step before convergence of the used MCMC algorithms are deleted, while thinning done in the second step should guarantee that the remaining samples can be considered as being independent. For each of the observed Bayesian models $50,000$ (dependent) samples are generated, except for the Bayesian adaptive lasso where $70,000$ ones are produced. This results in $4,000$ i.i.d. samples each, except of $6,000$ samples for the adaptive lasso. The reason for the special treatment is that in our experiments the Bayesian adaptive lasso turned out to require more samples to get stable estimates of the target variables.

\subsection{Real-world studies}
\label{realWorld}

In this section the performance of the proposed method is compared with the performance of other well-established methods based on three real-world datasets. The comparison is made along the lines of the above considerations (see beginning of Section \ref{exp}).

\subsubsection{The diabetes dataset}
\label{diabetes}
In this section the popular diabetes dataset, originally used by  \cite{efron2004} to examine the so-called LARS algorithm, is used for the performance evaluation. This dataset includes ten predictors, age,  sex,  body  mass  index,  average  blood  pressure,
and six blood serum measurements, measured from $n=442$ diabetes patients. The target variable of interest is a  quantitative  measure  of  disease progression  one  year  after  baseline. In the R-package \textit{care} (\cite{Rcare}) a standardized version (all variables have zero mean and variance equal to one) of the diabetes dataset is available. This version is used in our experiments.

The hyperparameters of the proposed Bayesian approach are specified as follows:  In an empirical Bayes manner, the parameters $p_1,...,p_p$ which belong to the prior of $\mathcal{A}$ and represent the a-priori belief in the importance of the predictors $x_1,...,x_p$ are assigned appropriate multiples (they must sum up to one) of the absolute values of the regression coefficients of the ridge regularized model which also determines the penalization factors of the adaptive lasso model. Moreover, the parameters $\tilde{p}_1,...,\tilde{p}_p$ (see Section \ref{metropolis}) are specified identically to the parameters $p_1,...,p_p$. As already mentioned in Section \ref{model}, one could define the mapping $\tilde{p}$ as the probability mass function of a zero truncated binomial distribution with size parameter equal to the number of predictors $p$ and the second parameter chosen in such a way that the mean of the distribution equals the a-priori most probable model size. Thus, a zero truncated binomial prior is assigned to the unknown model size $k$. However, in general it is somewhat cumbersome to determine the second parameter in such a way that the distribution has a given mean. Note that the expected value of a zero truncated binomial distribution with size parameter $p$ and second parameter $q$ equals $pq/(1 - (1-q)^p)$. Further, note that by the Abel-Ruffini theorem there is no algebraic solution to the general polynomial equations of degree five or higher with arbitrary coefficients. A classical binomial distribution with size parameter $p$ and second parameter equal to $\mu/p$ has mean $\mu$. Thus, a zero truncated binomial distribution with these parameters will result in a distribution with expected value not too far away from $\mu$. Therefore, defining $\tilde{p}$  as probability mass function of such a zero truncated distribution is a good alternative and our final choice in the performance evaluation of the diabetes dataset. Now the only question remaining is how to specify $\mu$. Since there is no a-priori knowledge available one could determine a useful value of $\mu$ via cross-validation in an empirical Bayes manner. However, to save time we empirically tried some possible values for $\mu$ and very soon ended up with the specification $\mu=9.9$ (which results in a good performance as one can see later on in this section). The parameters $a$ and $b$ are both set to $0.001$ which will also hold for the remaining experimental studies, as already reported in Section \ref{methodology}. Finally, the tuning parameters (see Section \ref{metropolis}) $p_h,\varepsilon_{\sigma}$ and $\varepsilon_{g}$ are respectively set to the values $0.4, 0.1$ and $60$.

Performing a $5$-fold cross-validation as described at the beginning of Section \ref{exp} results in Table \ref{tab:1}. The proposed approach achieves the lowest MMSE as well as the lowest MMAD and thus performs better than all methods under comparison.
\begin{table}
\caption{Diabetes data performance comparison: Median of mean squared prediction errors (MMSE) and median of mean absolute prediction deviations (MMAD) based on a $5$-fold cross-validation.}
\label{tab:1}       
\centering
\begin{tabular}{lll}
\hline\noalign{\smallskip}
Method & MMSE & MMAD \\
\noalign{\smallskip}\hline\noalign{\smallskip}
New approach & 0.4873534 & 0.5678801 \\
Lasso & 0.492067 & 0.571596 \\
Adaptive lasso & 0.4939229 & 0.5736721 \\
Elastic net & 0.4922686 & 0.5706994 \\
Bayesian lasso & 0.4924316 & 0.5736084 \\
Bayesian adaptive lasso & 0.4997307 & 0.5786672\\
Bayesian elastic net & 0.4895844 & 0.5727555\\
\noalign{\smallskip}\hline
\end{tabular}
\end{table}

\subsubsection{The prostate cancer dataset}
\label{prostate}

In this section the performance of the seven considered methods is compared on the prostate cancer data provided by \cite{Stamey1989}. This dataset examines the relationship between the level of prostate specific antigen (lpsa) and eight clinical measures from $99$ patients who were about to receive a radical prostatectomy. The eight measurments are log cancer volume (lcavol), log prostate weight (lweight), age, log benign prostatic hyperplasia amount (lbph), seminal vesicle invasion (svi), log capsular penetration (lcp), Gleason score (gleason), and percentage Gleason scores 4 or 5 (pgg45). This dataset is included e.g. in the R-package \textit{lasso2} (\cite{Rlasso2}). The predictors are standardized, consequently we only have to zero-center the target variable lpsa to make the data compatible with the model proposed by us.

The hyperparameters of our new Bayesian approach are specified as follows:
The parameters $p_1,...,p_p$ and $\tilde{p}_1,...,\tilde{p}_p$ are specified the same way as in Section \ref{diabetes}. Again, $\tilde{p}$ is defined as probability mass function of a zero truncated binomial distribution with size parameter $p$ and second parameter $\mu/p$, but now $\mu$ is specified as the number of nonzero regression coefficients from the corresponding lasso model. The remaining parameters are also defined as in Section \ref{diabetes}.

Thus, performing a $5$-fold cross-validation as described at the beginning of Section \ref{exp} results in Table \ref{tab:2}.
\begin{table}
\caption{Prostate cancer data performance comparison: Median of mean squared prediction errors (MMSE) and median of mean absolute prediction deviations (MMAD) based on a $5$-fold cross-validation.}
\label{tab:2}       
\centering
\begin{tabular}{lll}
\hline\noalign{\smallskip}
Method & MMSE & MMAD \\
\noalign{\smallskip}\hline\noalign{\smallskip}
New approach & 0.5068169 & 0.6089882 \\
Lasso & 0.5214503 & 0.6023388 \\
Adaptive lasso & 0.5429631 & 0.5990569 \\
Elastic net & 0.5293173 & 0.5837592 \\
Bayesian lasso & 0.5462163 & 0.5916963 \\
Bayesian adaptive lasso & 0.547966 & 0.5614767\\
Bayesian elastic net & 0.5267334 & 0.6138624\\
\noalign{\smallskip}\hline
\end{tabular}
\end{table}
Again, our method achieves the lowest MMSE, while some other methods go along with better MMADs on the prostate cancer dataset.

\subsubsection{The ozone dataset}
\label{ozone}

In this section the ozone datset originally studied by  \cite{Breiman1985} is used for the performance comparison. The data was measured in the area of Upland, CA, east of Los Angeles on 330 days in 1976. The target variable is the daily maximum one-hour-average ozone reading (parts per million) and the predictors are:
\begin{itemize}
\item Temperature (degrees F)
\item Inversion base height (feet)
\item Daggett pressure gradient (mmHg)
\item Visibility (miles)
\item Vandenburg 500 millibar height (m)
\item Humidity (\%)
\item Inversion base temperature (degrees F)
\item Wind speed (mph)
\end{itemize}
The dataset is available in the R-package \textit{gclus} (\cite{Rgclus}). We decide to examine the regression model including all linear, quadratic and two-way interactions, resulting in $44$ possible predictors. Moreover, we zero center and standardize (variance equal to one) all variables previous to the training of the diverse models. We want to mention that in contrast to a statement at the beginning of Section \ref{exp} in this performance evaluation for all Bayesian models $100,000$ MCMC samples are drawn and not $50,000$ or $70,000$ as stated there. This is the only experimental study where this statement is violated.

The hyperparameters of the newly proposed Bayesian approach are specified as follows:
We assign to each of the parameters $p_1,...,p_p$ and $\tilde{p}_1,...,\tilde{p}_p$ the value $1/p$, i.e. assume equal a-priori importance of the predictors. The function $\tilde{p}$ is defined in such a way, that it maps a value $k\in\left\{1,...,p\right\}$ to a value proportional to the third power of the probability $P(x = k)$, where $x$ denotes  a random variable which is zero truncated binomial distributed with size parameter $p$ and second parameter $7/p$. We empirically discover that taking the power of three can be a helpful step to guarantee fast convergence of the proposed MH algorithm (see Section \ref{metropolis}) in case of an increasing number of predictors and a prior based on a binomial distribution. By choosing the constant of proportionality in the definition of $\tilde{p}$ in such a way, that $\tilde{p}$ defines a probability distribution, the resulting distribution will have a smaller variance than the generating binomial distribution. Finally, the tuning parameters (see Section \ref{metropolis}) $p_h,\varepsilon_{\sigma}$ and $\varepsilon_{g}$ are  set to the values $0.5, 0.1$ and $60$, respectively.

Performing a $5$-fold cross-validation as described at the beginning of Section \ref{exp} results in Table \ref{tab:3}. The new approach achieves the lowest MMAD and a MSE comparable to the other methods .
\begin{table}
\caption{Ozone data performance comparison: Median of mean squared prediction errors (MMSE) and median of mean absolute prediction deviations (MMAD) based on a $5$-fold cross-validation.}
\label{tab:3}       
\centering
\begin{tabular}{lll}
\hline\noalign{\smallskip}
Method & MMSE & MMAD \\
\noalign{\smallskip}\hline\noalign{\smallskip}
New approach & 0.2471677 & 0.3837716 \\
Lasso & 0.2451734 & 0.3898133 \\
Adaptive lasso & 0.2539498 & 0.3893223 \\
Elastic net & 0.2498921 & 0.3898747 \\
Bayesian lasso & 0.256831 & 0.3926603 \\
Bayesian adaptive lasso & 0.2458003 & 0.3865291\\
Bayesian elastic net & 0.2409481 & 0.3853373\\
\noalign{\smallskip}\hline
\end{tabular}
\end{table}

\subsection{Simulation studies}
In this section, on the basis of simulated data corresponding to two different artificial models, the performance of our new method  is compared with the performance of other well-established methods. The artificial models on purpose include many correlated predictors, from which only a small subset is predictive, i.e. has regression coefficients different from zero. This should reflect difficult variable selection problems that more and more organizations have top face in their daily life nowadays and in the future. From both artificial models multiple datasets are simulated. In particular, $100$ datasets are sampled, each according to three different settings. First the sampling takes place with $n=50$ training observations, then with $n=100$ training observations and finally with $n=200$ training observations. The number of test observations is always the same and given by the value $n_{te}=200$. Moreover, all the observations are drawn independently from each other. For each sampled dataset the entire target vector (includes target values from both the training set and the testing set) is standardized via division by the sample standard deviation of the training target values. This simplifies finding a good choice of the tuning parameters for the MH algorithm proposed. The scaling allows for choosing them equal to specifications used in Section \ref{realWorld}. The predictors are not scaled at all since they are sampled from distributions with mean zero and standard deviation one.  

\subsubsection{Simulation study 1}
\label{sim1}
In this simulation study data are generated from the model
\begin{align}
y &= \beta_1x_1+...+\beta_{100}x_{100}+\varepsilon\\
\varepsilon &\sim\mathcal{N}(0,1)\\
(x_1,...,x_{100})^T &\sim\mathcal{N}(\mathbf{0},\mathbf{\Sigma})\\
\operatorname{diag}(\mathbf{\Sigma})&=\mathbf{1}\\
\Sigma_{i,j}&=0.6~\text{for}~i\neq j
\end{align}
with $(\beta_2,\beta_{11},\beta_{21},\beta_{51},\beta_{71},\beta_{81})=(-2.5,-2,-1.5,1.5,2,2.5)/\sqrt3$ and the remaining coefficients equal to zero.
This model is inspired by those used to evaluate the performance of the spike-and-slab lasso by \cite{Ro2018}.

The hyperparameters of the proposed Bayesian approach are specified as follows:
The parameters $p_1,...,p_p$ are defined as multiples of the absolute sample correlations between the training responses $\mathbf{y}$ and the training predictors $\mathbf{x}_1,...,\mathbf{x}_p$. Note, that the multiplicative factor has to be defined in such a way that the parameters sum up to one. Moreover, the parameters $\tilde{p}_1,...,\tilde{p}_p$ are specified identical to the parameters $p_1,...,p_p$. The function $\tilde{p}$ maps each value $k\in\left\{1,...,p\right\}$ to a value proportional to the third power of the probability $P(x = k)$, where $x$ denotes  a random variable which is zero truncated binomially distributed with size parameter $p$ and second parameter $2.5/p$. Finally, the tuning parameters $p_h,\varepsilon_{\sigma}$ and $\varepsilon_{g}$ are  set to the values $0.5, 0.1$ and $60$, respectively.

\begin{figure*}
\centering
\begin{minipage}{.45\textwidth}
  \centering
  \includegraphics[width=\linewidth]{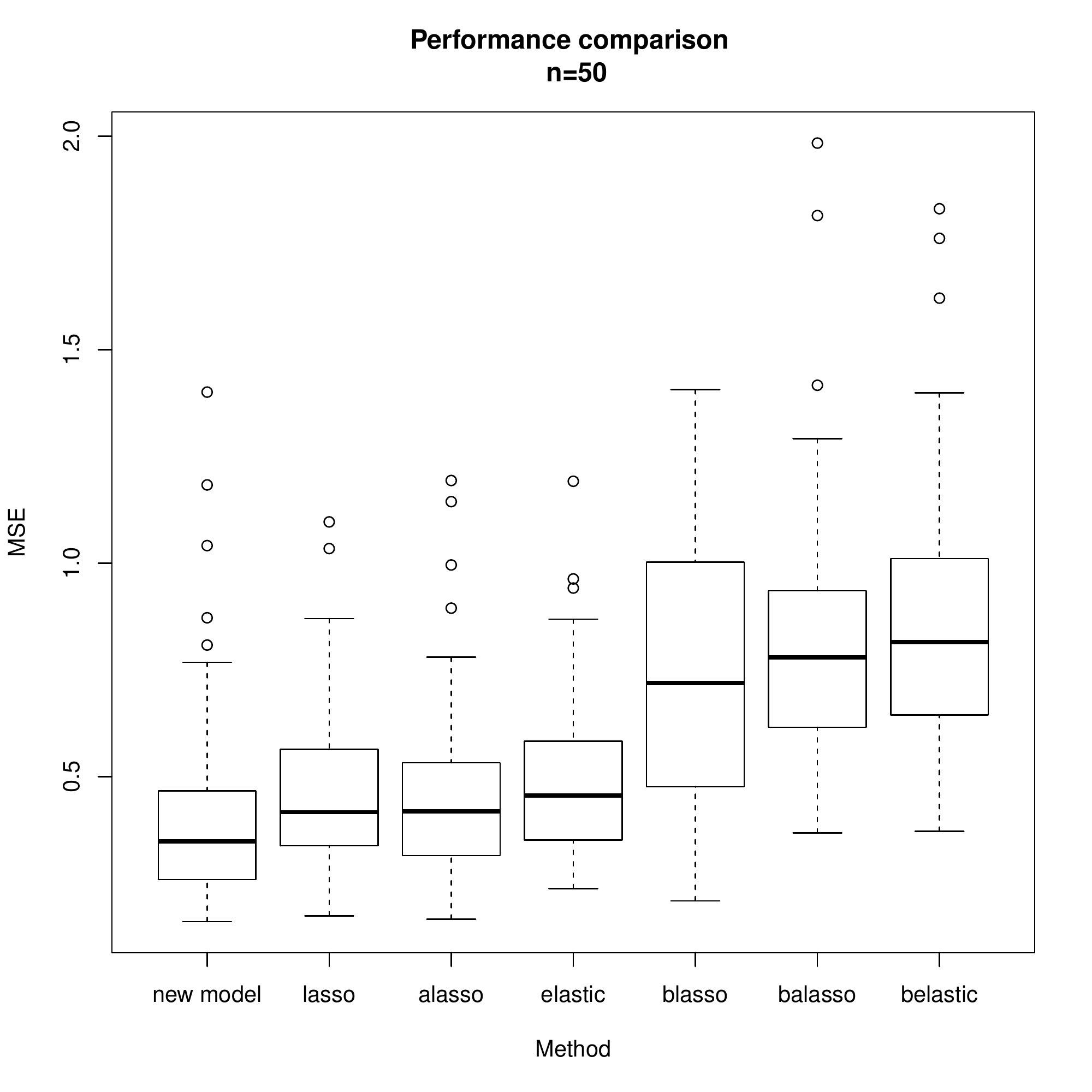}
  \caption{Boxplots of the MSEs obtained from the $100$ simulated datasets with $n=50$ training samples.}
  \label{MSE50}
\end{minipage}%
\hspace{.09\textwidth}
\begin{minipage}{.45\textwidth}
  \centering
  \includegraphics[width=\linewidth]{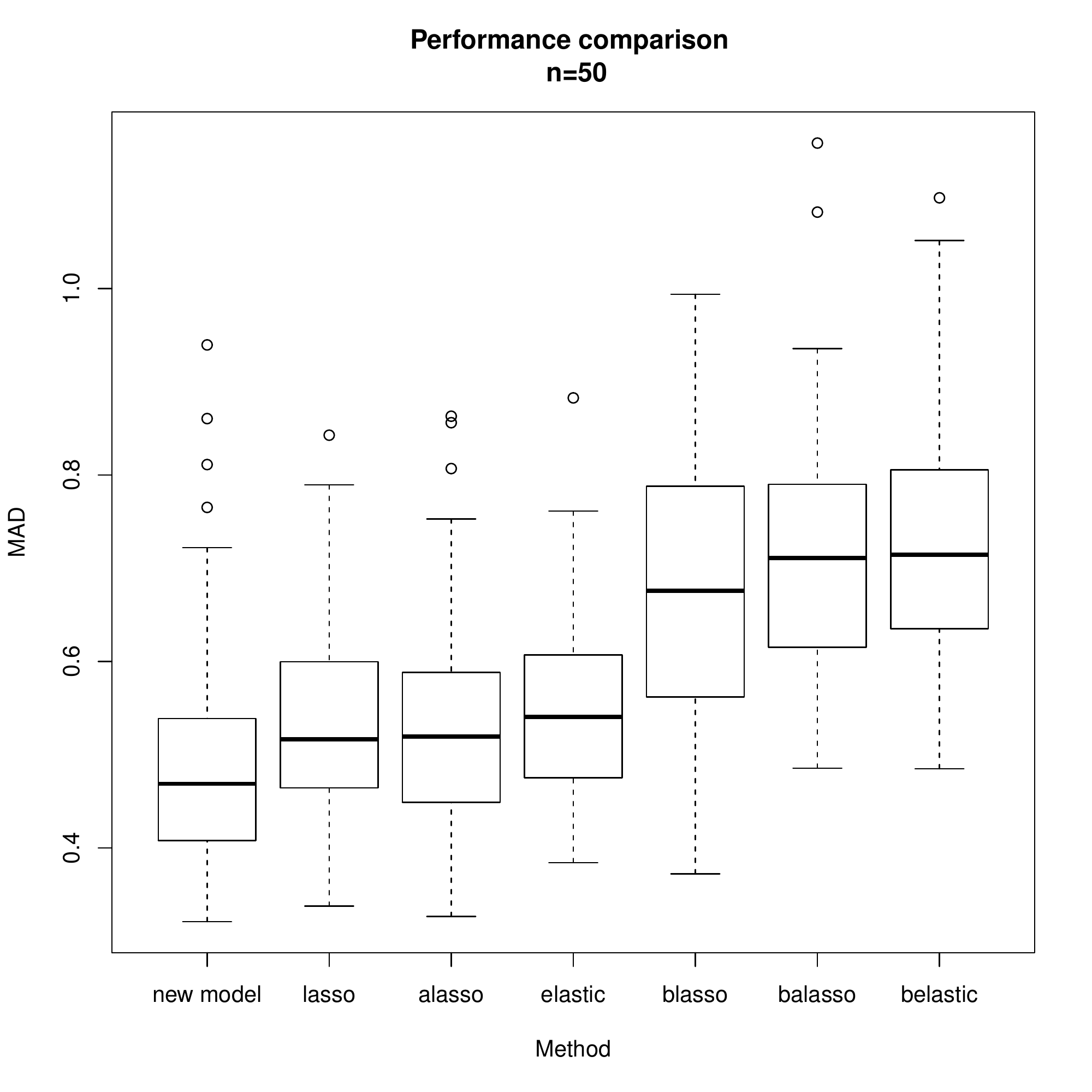}
  \caption{Boxplots of the MADs obtained from the $100$ simulated datasets with $n=50$ training samples.}
  \label{MAD50}
\end{minipage}  
\\
  \begin{minipage}{.45\textwidth}
  \centering
  \includegraphics[width=\linewidth]{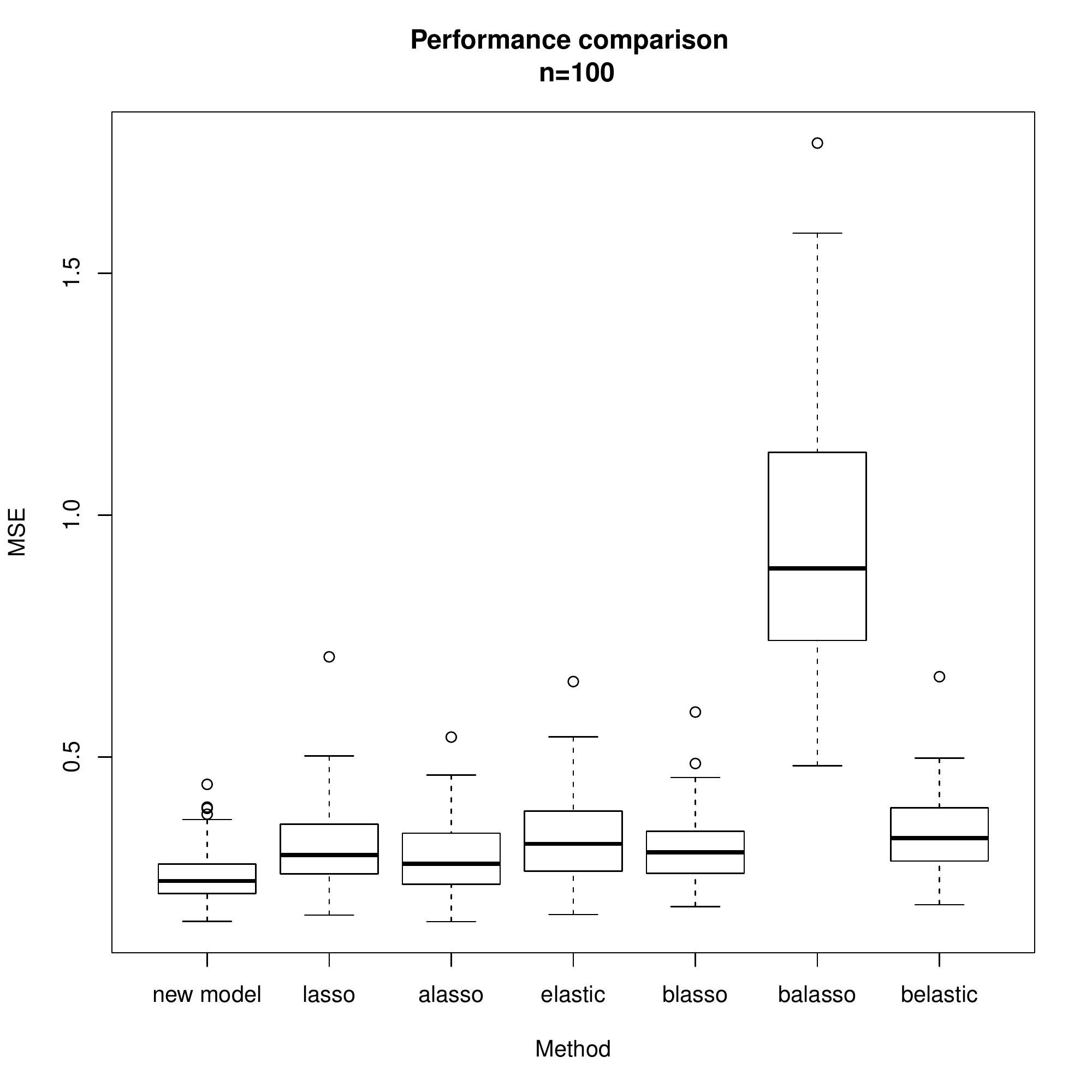}
  \caption{Boxplots of the MSEs obtained from the $100$ simulated datasets with $n=100$ training samples.}
  \label{MSE100}
\end{minipage}%
\hspace{.09\textwidth}
\begin{minipage}{.45\textwidth}
  \centering
  \includegraphics[width=\linewidth]{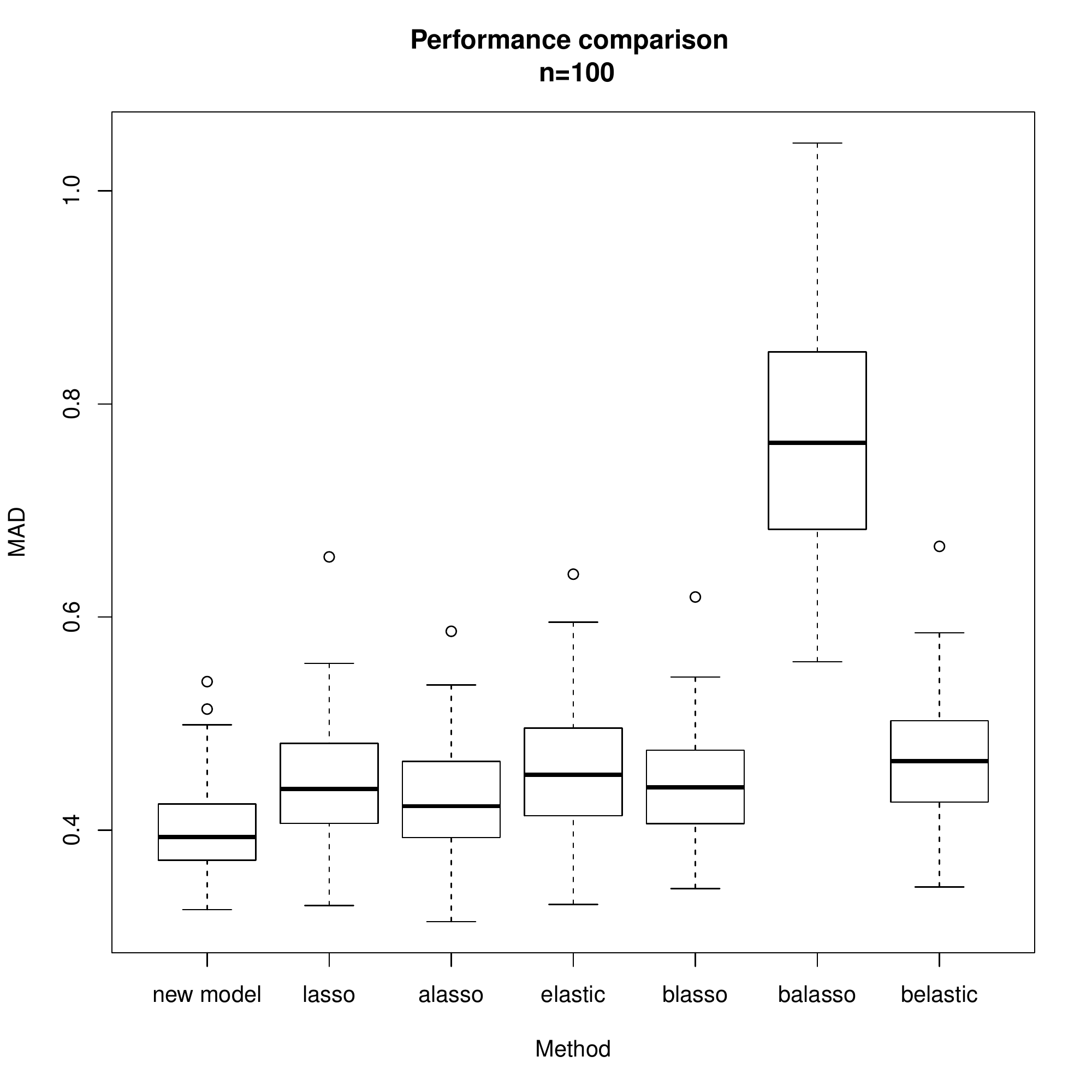}
  \caption{Boxplots of the MADs obtained from the $100$ simulated datasets with $n=100$ training samples.}
  \label{MAD100}
\end{minipage}    
\end{figure*}

Aggregating the accuracy measures obtained by training the models to be compared, as described at the beginning of Section \ref{exp}, results in Table \ref{tab:4}. The proposed approach achieves the lowest MMSE as well as the lowest MMAD and thus performs better than all other methods.
\begin{table}
\caption{Performance comparison for different specifications of $n$: Median of mean squared prediction errors (MMSE) and median of mean absolute prediction deviations (MMAD) based on a $100$ simulated datasets.}
\label{tab:4}       
\centering
\begin{tabular}{lllllll}
\hline\noalign{\smallskip}
Method & MMSE  & MMAD & MMSE &  MMAD & MMSE & MMAD\\
&$n=50$&$n=50$&$n=100$&$n=100$&$n=200$&$n=200$\\
\noalign{\smallskip}\hline\noalign{\smallskip}
New approach            & 0.3390823 & 0.4618555    & 0.2436617 & 0.3933677      & 0.234171  & 0.3868592 \\
Lasso                   & 0.4128203 & 0.5142703    & 0.2972025 & 0.4385642      & 0.2612206 & 0.4064837 \\
Adaptive lasso          & 0.4146856 & 0.5174635    & 0.2796941 & 0.422432       & 0.2606342 & 0.4076061 \\
Elastic net             & 0.4557527 & 0.5405664    & 0.3204951 & 0.4521248      & 0.2667704 & 0.4108516 \\
Bayesian lasso          & 0.7189461 & 0.6758523    & 0.3032926 & 0.439959       & 0.2596804 & 0.4053971 \\
Bayesian adaptive lasso & 0.7795937 & 0.7110935    & 0.8896619 & 0.7637107      & 0.3793791 & 0.4924849\\
Bayesian elastic net    & 0.8148157 & 0.7146019    & 0.3324611 & 0.4647419      & 0.2668035 & 0.4122355\\
\noalign{\smallskip}\hline
\end{tabular}
\end{table}
In the Figures \ref{MSE50}-\ref{MAD100} boxplots of the MSEs and MADs obtained from the $100$ simulated datasets with $n=50$ and $n=100$ training observations can be found. These plots again illustrate that our new model performs significantly better than the other models considered. Moreover, one can see that the implementation of the Bayesian adaptive lasso has some serious problems with these simulated data, due to numerical instabilities.

\begin{figure*}
\centering
\begin{minipage}{.45\textwidth}
  \centering
  \includegraphics[width=\linewidth]{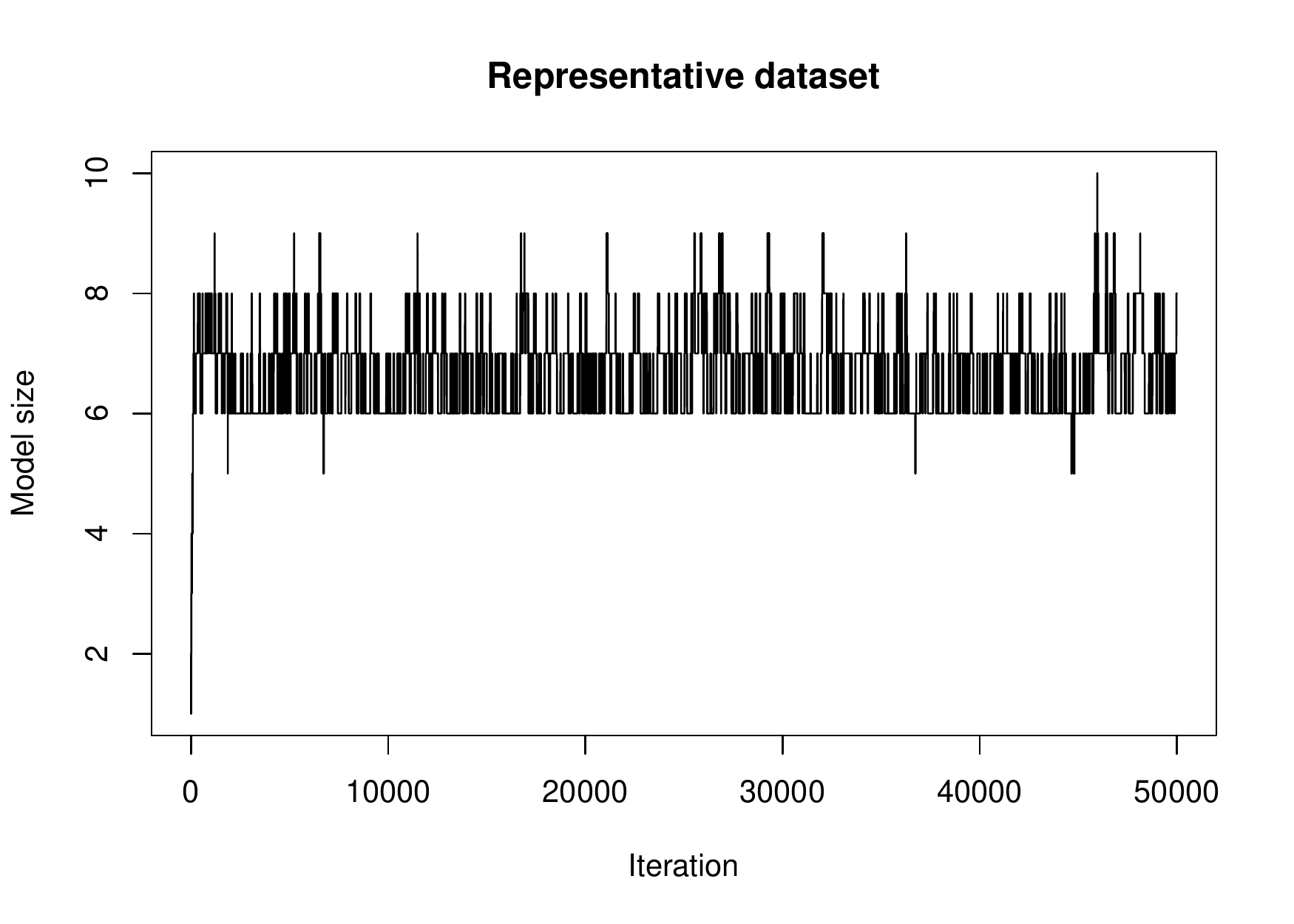}
  \caption{Markov chain of the model size for a representative dataset simulated from model $(61-65)$ with $100$ observations.}
  \label{size100}
\end{minipage}%
\hspace{0.09\textwidth}
\begin{minipage}{.45\textwidth}
  \centering
  \includegraphics[width=\linewidth]{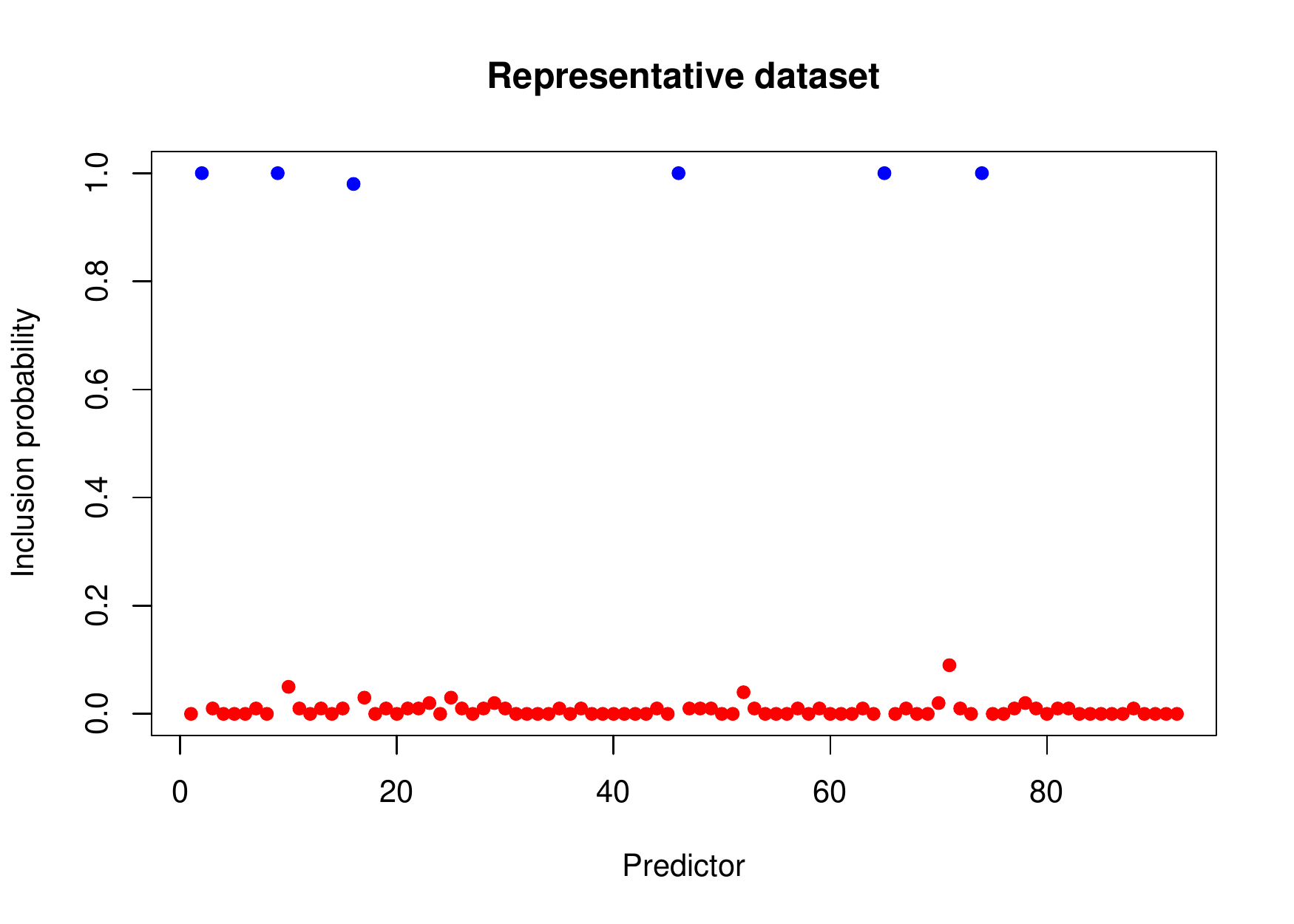}
  \caption{Posterior model inclusion probabilities of the predictors. Truly non-zero regression coefficients are colored blue, while the remaining coefficients are colored red.}
  \label{inclusion100}
\end{minipage}  
\end{figure*}

To provide some more insights into the new method proposed we consider a representative dataset simulated from model $(61-65)$ with $n=100$ observations. In Figure \ref{size100} one can see the Markov chain of the model size $k$ (i.e. the number of non-zero regression coefficients) produced from our MH algorithm. Further, in Figure \ref{inclusion100} the relative frequencies of a given regression coefficient being non-zero are plotted for the reduced Markov chain. By the latter one we mean the Markov chain remaining after the deletion step performed to obtain i.i.d. samples. The relative frequencies can be interpreted as the a posteriori probabilities of the predictors being included in the model. All the truly non-zero regression coefficients are colored blue, while the remaining coefficients are colored red. One can see that the proposed approach selects the true data generating model pretty well. Finally, we want to report that the acceptance rate of the  proposed MH algorithm is given by the value $0.20672$, which is a good acceptance rate indicating that the algorithm converges fast.

\subsubsection{Simulation study 2}
\label{sim2}

In this simulation study data are generated from the model
\begin{align}
y &= \beta_1x_1+...+\beta_{500}x_{500}+\varepsilon\\
\varepsilon &\sim\mathcal{N}(0,1)\\
(x_1,...,x_{500})^T &\sim\mathcal{N}(\mathbf{0},\mathbf{\Sigma})\\
\operatorname{diag}(\mathbf{\Sigma})&=\mathbf{1}\\
\Sigma_{i,j}&=0.75~\text{for}~i\neq j
\end{align}
with $(\beta_2,\beta_{11},\beta_{21},\beta_{51},\beta_{71},\beta_{81})=(-2.5,-2,-1.5,1.5,2,2.5)/\sqrt3$ and the remaining coefficients equal to zero. Simulating from this model results in more difficult datasets in context of variable selection than simulating from the model defined in Section \ref{sim1}. There are $400$ additional potential predictors and the pairwise correlations between them are increased from $0.6$ to $0.75$.

The hyperparameters of the newly proposed Bayesian approach are specified, in an empirical Bayes manner, as follows:
The parameters $p_1,...,p_p$ are assigned appropriate multiples (they must sum up to one) of the absolute values of the regression coefficients of the ridge regularized model, which also determines the penalization factors of the adaptive lasso model. The parameters $\tilde{p}_1,...,\tilde{p}_p$ are chosen analogously, except that the third power of the ridge coefficients is chosen. The function $\tilde{p}$ maps each value $k\in\left\{1,...,p\right\}$ to a value proportional to the third power of the probability $P(x = k)$, where $x$ denotes  a random variable which is zero truncated binomially distributed with size parameter $p$ and second parameter $1.75/p$. Finally, the tuning parameters $p_h,\varepsilon_{\sigma}$ and $\varepsilon_{g}$ are  set to the values $0.5, 0.1$ and $60$, respectively.

\begin{figure*}
\centering
\begin{minipage}{.45\textwidth}
  \centering
  \includegraphics[width=\linewidth]{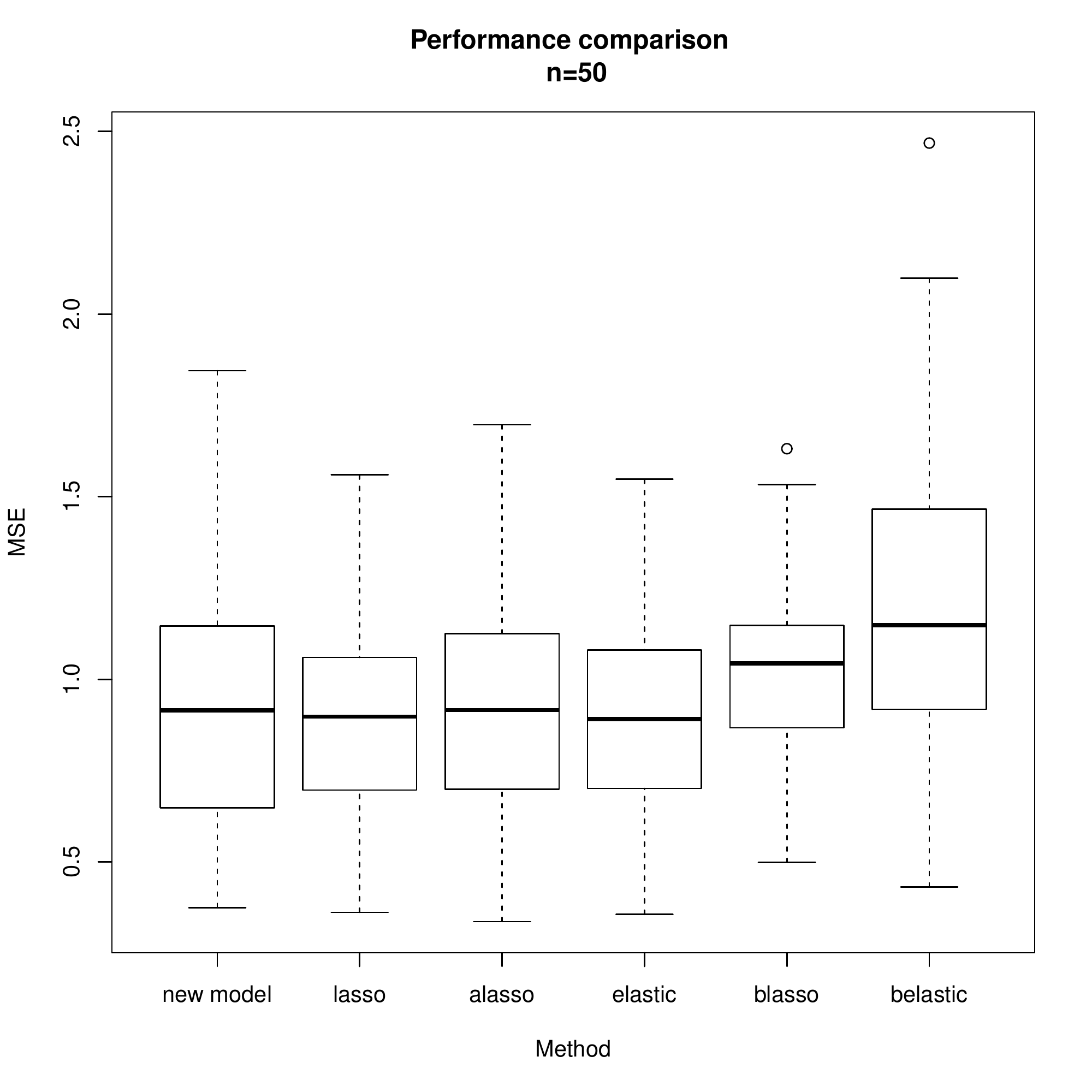}
  \caption{Boxplots of the MSEs obtained from the $100$ simulated datasets with $n=50$ training samples.}
  \label{2MSE50}
\end{minipage}%
\hspace{.09\textwidth}
\begin{minipage}{.45\textwidth}
  \centering
  \includegraphics[width=\linewidth]{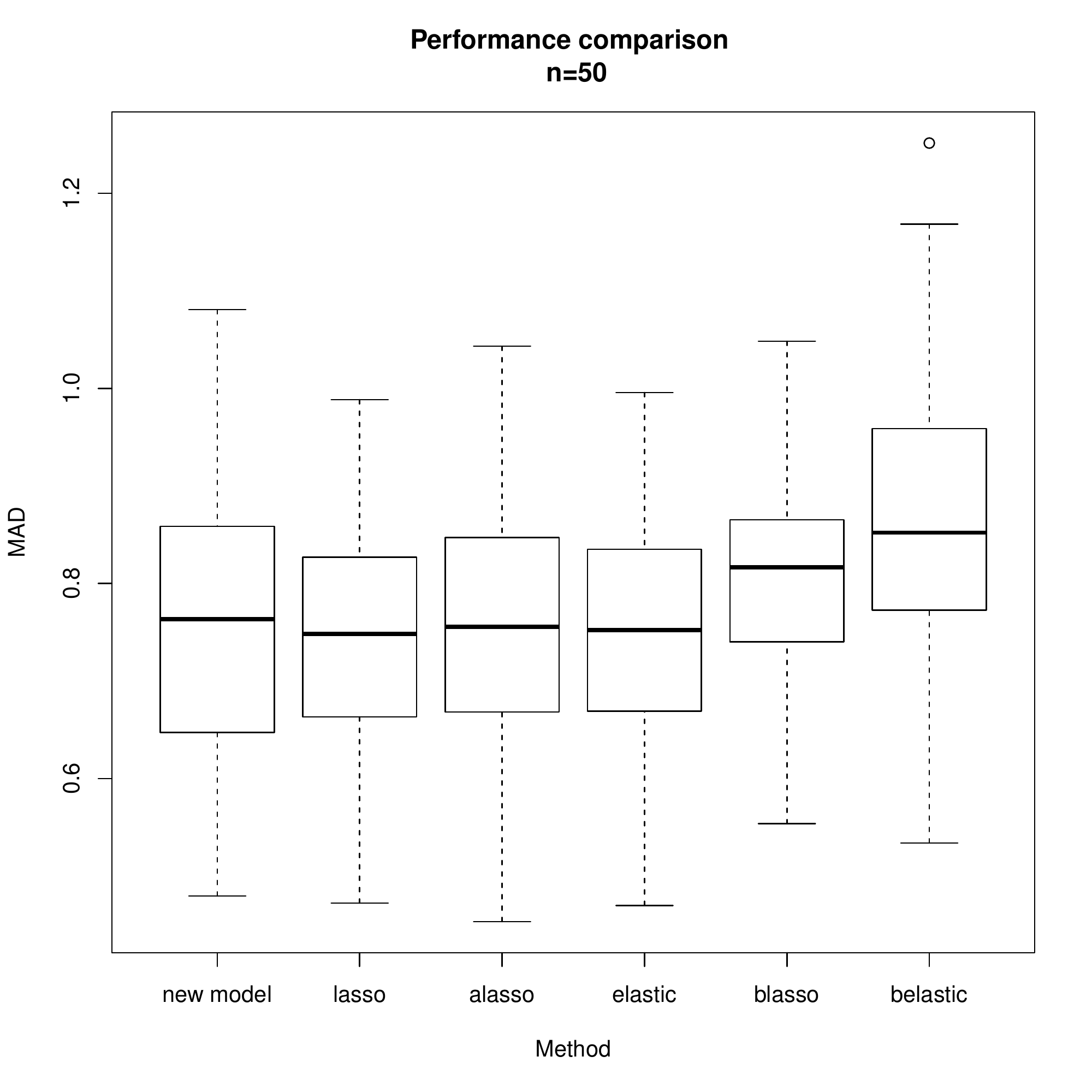}
  \caption{Boxplots of the MADs obtained from the $100$ simulated datasets with $n=50$ training samples.}
  \label{2MAD50}
\end{minipage}  
\\
  \begin{minipage}{.45\textwidth}
  \centering
  \includegraphics[width=\linewidth]{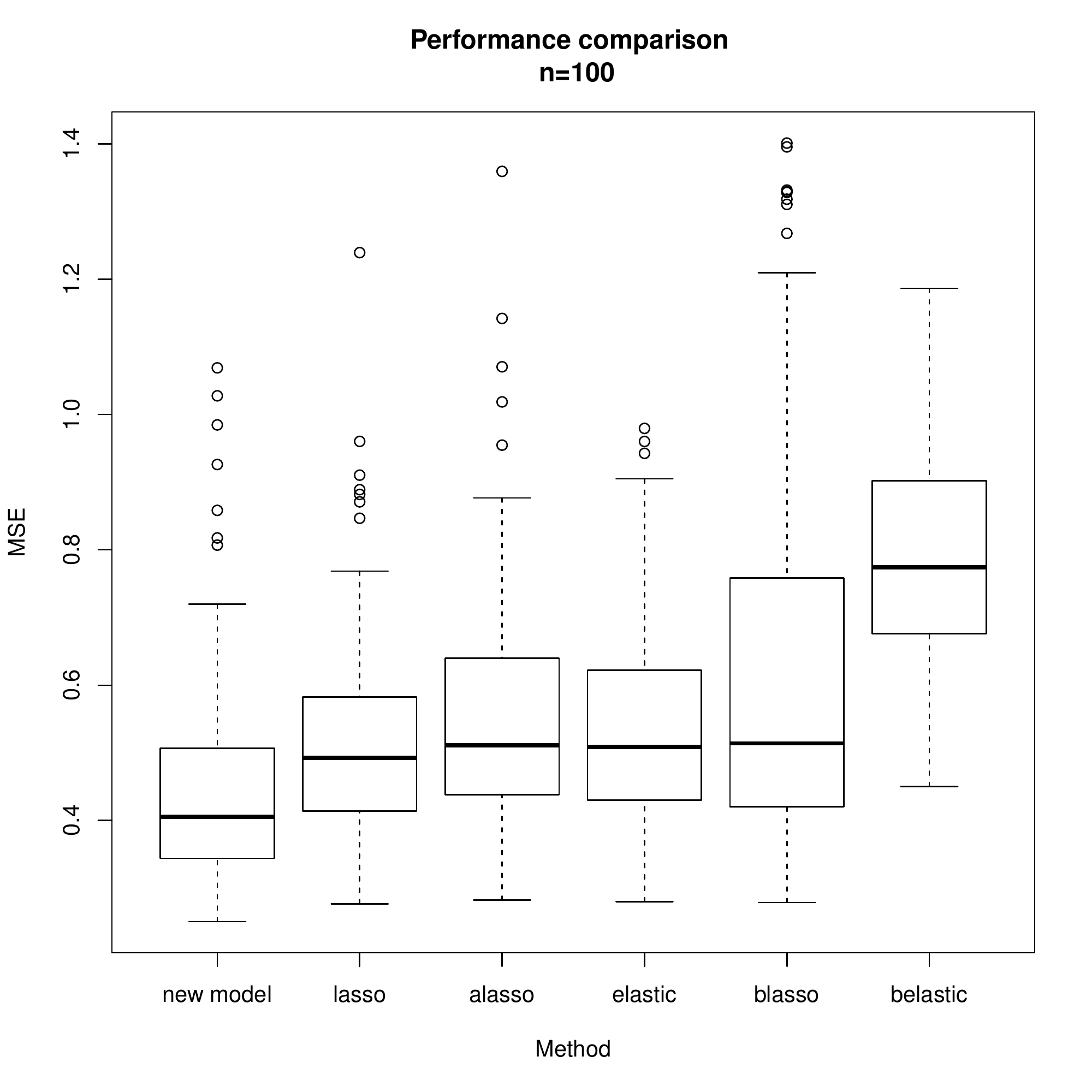}
  \caption{Boxplots of the MSEs obtained from the $100$ simulated datasets with $n=100$ training samples.}
  \label{2MSE100}
\end{minipage}%
\hspace{.09\textwidth}
\begin{minipage}{.45\textwidth}
  \centering
  \includegraphics[width=\linewidth]{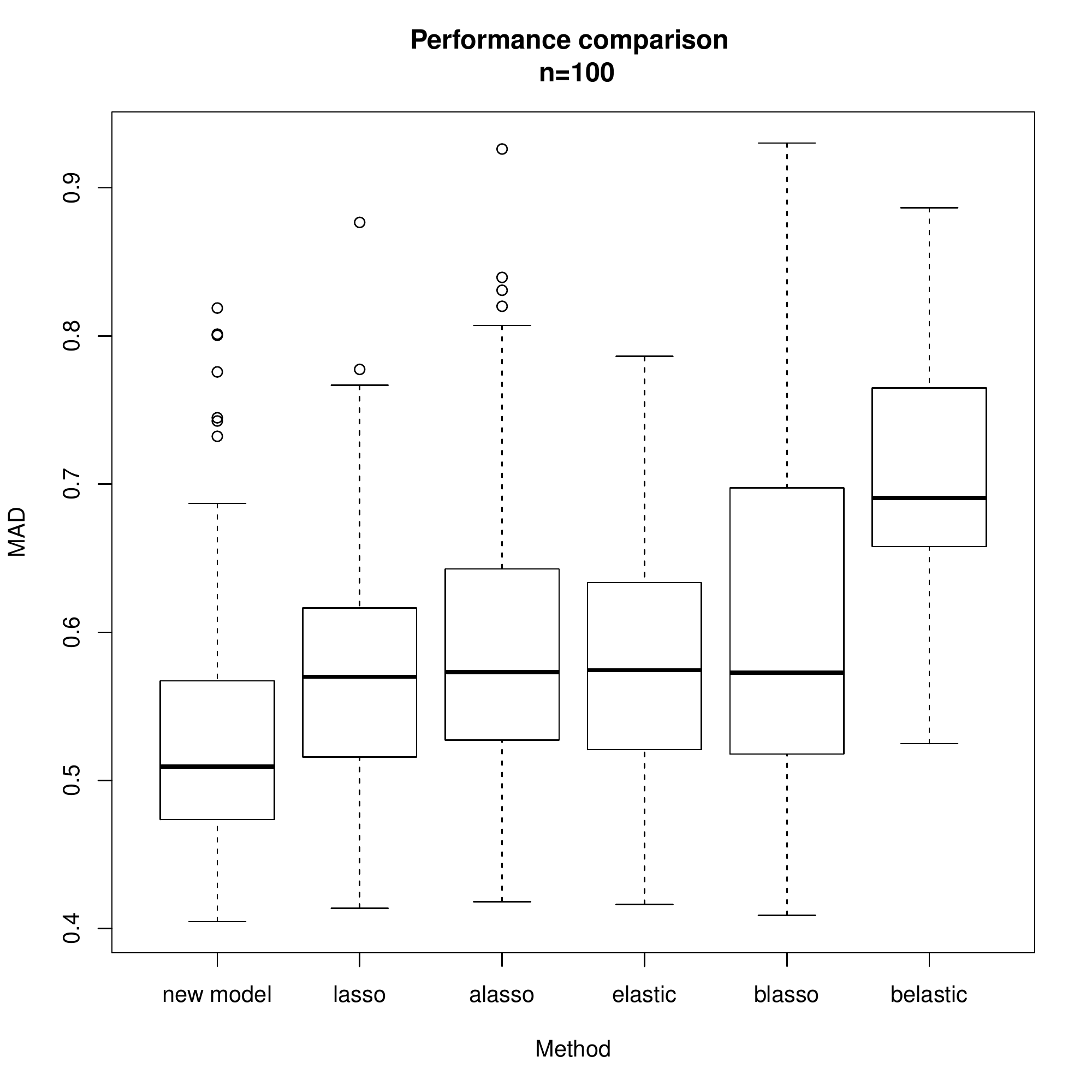}
  \caption{Boxplots of the MADs obtained from the $100$ simulated datasets with $n=100$ training samples.}
  \label{2MAD100}
\end{minipage}    
\end{figure*}

Due to numerical issues in the corresponding R-implementation the Bayesian adaptive lasso has to be excluded from the model comparison. Aggregating the accuracy measures obtained by training the remaining models results in Table \ref{tab:5}. The proposed approach achieves the lowest MMSE as well as the lowest MMAD in the settings with $n=100$ and $n=200$ training observations. In the setting with $n=50$ training observations it achieves a comparable accuracy. The Figures \ref{2MSE50}-\ref{2MAD100} show boxplots of the MSEs and MADs obtained from the $100$ simulated datasets with $n=50$ and $n=100$ training observations. In the setting $n=100$ our method outperforms the other ones. The median of the MADs as well as the median of the MSEs lies below all corresponding boxes of the other methods. 

\begin{table}
\caption{Performance comparison for different specifications of $n$: Median of mean squared prediction errors (MMSE) and median of mean absolute prediction deviations (MMAD) based on a $100$ simulated datasets.}
\label{tab:5}       
\centering
\begin{tabular}{lllllll}
\hline\noalign{\smallskip}
Method & MMSE  & MMAD & MMSE &  MMAD & MMSE & MMAD\\
&$n=50$&$n=50$&$n=100$&$n=100$&$n=200$&$n=200$\\
\noalign{\smallskip}\hline\noalign{\smallskip}
New approach            & 0.9155098 & 0.7634292    & 0.4057127 & 0.5094444      & 0.343676   & 0.4678344 \\
Lasso                   & 0.8979915 & 0.7481239    & 0.4926306 & 0.5699708      & 0.4065196  & 0.5115676 \\
Adaptive lasso          & 0.9159057 & 0.7556247    & 0.5108755 & 0.5731952      & 0.4093501  & 0.5127194 \\
Elastic net             & 0.8918046 & 0.7524384    & 0.5090338 & 0.574375       & 0.4080551  & 0.5138157 \\
Bayesian lasso          & 1.043241  & 0.8165255    & 0.5140319 & 0.572721       & 0.4343926  & 0.5303905 \\
Bayesian elastic net    & 1.147992  & 0.8520899    & 0.7743223 & 0.6907885      & 0.8825549  & 0.7553813 \\
\noalign{\smallskip}\hline
\end{tabular}
\end{table}

\section{Conclusion}
\label{conclusion}

In this article we have presented a novel Bayesian approach to cope with the problem of variable selection in the multiple linear regression model with dependent predictors. To make our method robust to challenges such as multicollinearity, or the number of observations being smaller than the number of potential predictors, a $g$-prior with an additional ridge parameter was assigned to the unknown regression coefficients. While other authors apply Gibbs sampling to simulate from the joint posterior distribution we presented an intelligent Metropolis Hastings algorithm for this task. Thus, we do not need to compute the conditional distributions needed for the Gibbs sampling algorithm, which often are not available in an analytically closed form, and have to be approximated. Further, we have shown that the proposed Bayesian approach is consistent in terms of model selection under some nominal assumptions. Experimental studies with three different real-world datasets as well as simulated datasets from two artificial models demonstrated the good performance of the presented method. In particular, on the simulated datasets our new approach performed significantly better than all other well-established methods.

\section*{Acknowledgments}
The publication is one of the results of the project iDev40 (www.idev40.eu). The iDev40 project  has  received  funding  from  the  ECSEL  Joint  Undertaking (JU) under  grant  agreement  No  783163. The JU receives support from the European Union’s Horizon  2020  research  and  innovation  programme. It  is  co-funded  by  the  consortium  members, grants from Austria, Germany, Belgium, Italy, Spain and Romania.
\vspace{-2cm}
\begin{figure}[H]
\centering
\begin{minipage}{.3\textwidth}
  \centering
  \includegraphics[width=\linewidth]{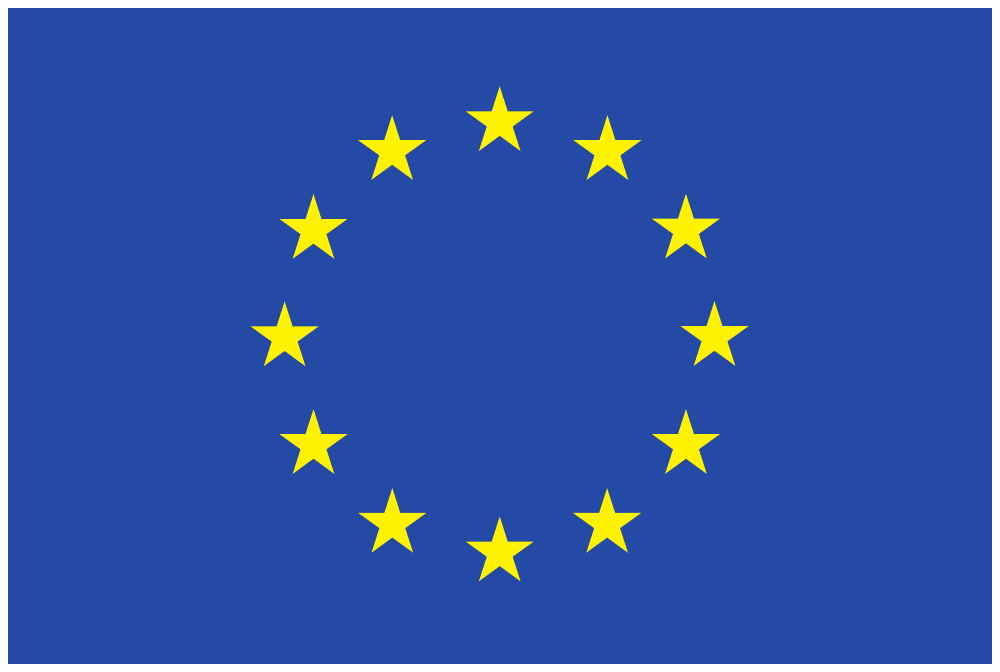}
\end{minipage}%
\hspace{.01\textwidth}
\begin{minipage}{.3\textwidth}
  \centering
  \includegraphics[width=\linewidth]{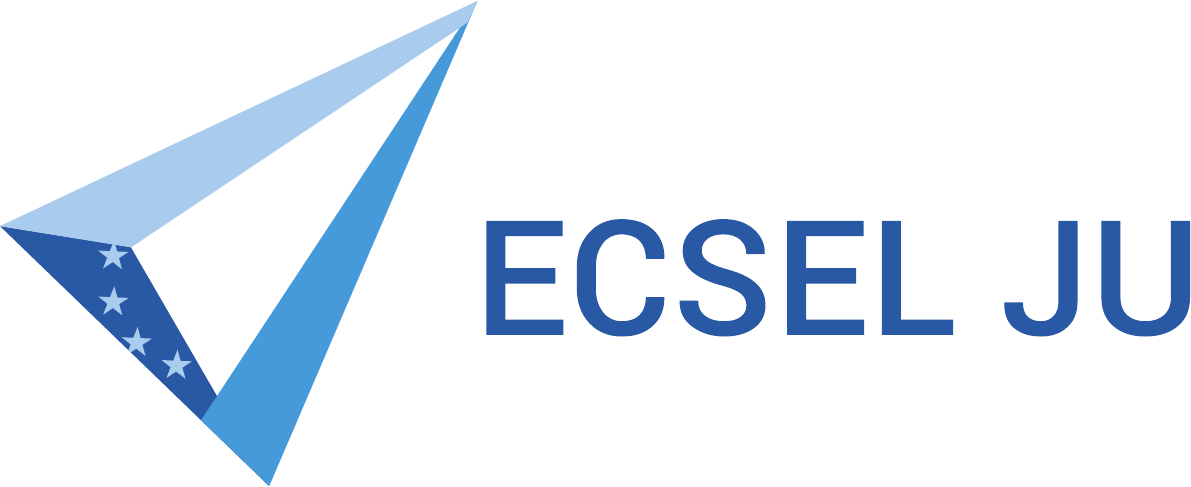}
\end{minipage} 
\hspace{.07\textwidth}
\begin{minipage}{.3\textwidth}
  \centering
  \includegraphics[width=\linewidth]{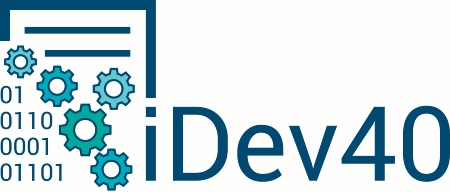}
\end{minipage}    
\end{figure}

\vspace{-2cm}
\appendix

\section{Proof of Theorem 1}
\label{appendix}

Note that for showing $(14)$ it suffices to show that
\begin{equation}
\underset{n\rightarrow \infty}{\operatorname{p~lim~}} \frac{p(M_{\mathcal{A}^{\prime}}|\mathbf{y},\mathbf{X})}{p(M_{\mathcal{A}}|\mathbf{y},\mathbf{X})} = \underset{n\rightarrow \infty}{\operatorname{p~lim~}} \frac{p(\mathcal{A}^{\prime}|\mathbf{y},\mathbf{X})}{p(\mathcal{A}|\mathbf{y},\mathbf{X})} = 0
\end{equation}
holds, where $M_{\mathcal{A}}$ denotes the true model and $M_{\mathcal{A}^{\prime}}$ denotes another model different from the true one. The sets $\mathcal{A}$ and $\mathcal{A}^{\prime}$, respectively, denote the index sets of the active predictors (predictors probably being different from zero) of the two models  $M_{\mathcal{A}}, M_{\mathcal{A}^{\prime}}$ and thus uniquely determine them. In a first step of the proof we compute the joint posterior of $\mathcal{A}$ and $g$ up to a constant of proportionality. Simple linear algebra and the facts that the densities of multivariate normal distributions and inverse gamma distributions integrate to one reveal for $n>\zeta$:
\begin{equation}
p(\mathcal{A},g|\mathbf{y},\mathbf{X})\propto p(\mathcal{A})g^{-\frac{3}{2}}\operatorname{exp}\left(-\frac{n}{2g}\right)(g+1)^{-\frac{k}{2}+a+\frac{n}{2}}\left[1+g(1-R_{\mathcal{A}}^2)\right]^{-a-\frac{n}{2}}
\end{equation}
where
\begin{equation}
R_{\mathcal{A}}^2 = \frac{\mathbf{y}^TH_{\mathcal{A}}\mathbf{y}}{\mathbf{y}^T\mathbf{y}+2b},
\end{equation}
\begin{equation}
H_{\mathcal{A}} = \mathbf{X}_{\mathcal{A}}\left(\mathbf{X}_{\mathcal{A}}^T\mathbf{X}_{\mathcal{A}}\right)^{-1}\mathbf{X}_{\mathcal{A}}^T
\end{equation}
and $k$ denotes the cardinality of $\mathcal{A}$. In principle, the required marginal posterior of $\mathcal{A}$ can be found by integrating out $g$ in equation (A.2). However, there exists no closed-form solution of the corresponding integral. To overcome this problem, a lower and an upper bound of the marginal posterior are derived. The numerator of the quotient in (A.1) is then replaced by the upper bound and the denominator by the lower bound such that the resulting quotient is an upper bound of the original one. Showing that the upper bound of the original quotient converges to zero will immediately imply that (A.1) holds. To determine these bounds, at first, a property of the Gaussian hypergeometric function given by
\begin{equation}
_2F_1(\alpha,\beta,\tau,z)=\frac{\operatorname\Gamma(\tau)}{\operatorname\Gamma(\beta)\operatorname\Gamma(\tau-\beta)}\int\limits_{0}^{1}t^{\beta -1}(1-t)^{\tau-\beta-1}(1-tz)^{-\alpha}~dt
\end{equation}
and convergent for $|z|<1$ with $\tau>\beta>0$ and for $|z|=1$ only if $\tau > \alpha + \beta$ and $\beta>0$
is derived. It can be easily validated that for $1>z\in\mathbb{R}$ and $\tau>\beta>0$  the following equation holds:
\begin{align*}
&\int\limits_{0}^{\infty}t^{\beta -1}(1+t)^{\alpha - \tau}\left[1+t(1-z)\right]^{-\alpha}~dt\\
&=\frac{\operatorname{\Gamma(\beta)}\operatorname\Gamma(-\alpha-\beta+\tau)}{\operatorname\Gamma(-\alpha+\tau)}{_2F_1}(\alpha,\beta,1+\alpha+\beta-\tau,1-z)\\
&+\frac{\operatorname\Gamma(\alpha+\beta-\tau)\operatorname\Gamma(-\beta+\tau)}{\operatorname\Gamma(\alpha)} {_2F_1}(-\alpha+\tau,-\beta+\tau, 1-\alpha-\beta+\tau,1-z)(1-z)^{-\alpha-\beta+\tau}\\
&=\frac{\operatorname\Gamma(\beta)\operatorname\Gamma(\tau-\beta)}{\operatorname\Gamma(\tau)} {_2F_1}(\alpha,\beta,\tau,z)
\end{align*}
Note that the second equality holds by identity $15.3.6$ in  \cite{abramowitz1964}. Above equation directly translates to the following identity which will later on be used to evaluate an upper and a lower bound of $p(\mathcal{A}|\mathbf{y},\mathbf{X})$:
\begin{equation}
\int\limits_{0}^{\infty}t^{\beta}(1+t)^{\tau}[1+t(1-z)]^{\alpha}~dt=\frac{\operatorname\Gamma(\beta+1)\operatorname\Gamma(-\alpha-\tau-\beta-1)}{\operatorname\Gamma(-\alpha-\tau)}   {_2F_1}(-\alpha,\beta+1,-\alpha-\tau,z)
\end{equation}

Determination of the upper bound:\\
At first we observe that:
\begin{equation}
(g + 1)^{\frac{n}{2}-\frac{k}{2}+a}\leq g^{\frac{n}{2}-\frac{k}{2}+a}\operatorname{exp}\left(-\frac{n}{2g}\right)^{-1+\frac{k-2a}{n}}~\forall g\in\mathbb{R}^{+}
\end{equation}
This inequality holds since it can be reduced to the well known inequality $1+x\leq\operatorname{exp}(x)~\forall x\in\mathbb{R}$. Moreover, we observe that:
\begin{equation}
\operatorname{exp}\left(-\frac{n}{2g}\right)^{\frac{k-2a}{n}}\leq 1 ~\forall g\in\mathbb{R}^{+}
\end{equation}
Using the inequalities (A.7) and (A.8) an upper bound of $p(\mathcal{A}|\mathbf{y},\mathbf{X})$ can be derived:
\begin{align}
&p(\mathcal{A}|\mathbf{y},\mathbf{X})/p(\mathcal{A})\\
&\leq \int\limits_{0}^{\infty}g^{-\frac{3}{2}}\operatorname{exp}\left(-\frac{n}{2g}\right)^{\frac{k-2a}{n}}g^{\frac{n}{2}-\frac{k}{2}+a}\left[1+g(1-R_{\mathcal{A}}^2)\right]^{-a-\frac{n}{2}}~dg\\
&\leq\int\limits_{0}^{\infty}g^{\frac{n}{2}-\frac{k}{2}+a-\frac{3}{2}}\left[1+g(1-R_{\mathcal{A}}^2)\right]^{-a-\frac{n}{2}}~dg\\
&=\frac{\operatorname\Gamma(-\frac{k}{2}+a+\frac{n}{2}-\frac{1}{2})\operatorname\Gamma(\frac{k}{2}+\frac{1}{2})}{\operatorname\Gamma(a+\frac{n}{2})}{_2F_1}(a+\frac{n}{2},-\frac{k}{2}+a+\frac{n}{2}-\frac{1}{2},a+\frac{n}{2},R_{\mathcal{A}}^2)\\
&=\frac{\operatorname\Gamma(-\frac{k}{2}+a+\frac{n}{2}-\frac{1}{2})\operatorname\Gamma(\frac{k}{2}+\frac{1}{2})}{\operatorname\Gamma(a+\frac{n}{2})}{_2F_1}(0,\frac{k}{2}+\frac{1}{2},a+\frac{n}{2},R_{\mathcal{A}}^2)(1-R_{\mathcal{A}}^2)^{\frac{k}{2}-a-\frac{n}{2}+\frac{1}{2}}\\
&=\frac{\operatorname\Gamma(-\frac{k}{2}+a+\frac{n}{2}-\frac{1}{2})\operatorname\Gamma(\frac{k}{2}+\frac{1}{2})}{\operatorname\Gamma(a+\frac{n}{2})}(1-R_{\mathcal{A}}^2)^{\frac{k}{2}-a-\frac{n}{2}+\frac{1}{2}}
\end{align}
Note that (A.12) is a direct application of (A.6) to (A.11). Equality (A.13) holds by identity $15.3.3$ in  \cite{abramowitz1964}. The last equality holds, since $2F_1(0,\beta,\tau,z)$ equals the area under the density of a beta distribution with parameters $\beta$ and $\tau-\beta$ and thus is given by the value $1$. Applying Stirling's formula
$$\operatorname{\Gamma}(c_1x+c_2)\approx\sqrt{2\pi}\operatorname{exp}(-c_1x)(c_1x)^{c_1x+c_2-\frac{1}{2}}$$
to (A.14), which was also used in the model consistency proof of  \cite{Wang2015}, allows another approximate simplification of the right-hand side expression of (A.14) for large $n$ through:
\begin{equation}
\left(\frac{n}{2}\right)^{-\frac{k}{2}-\frac{1}{2}}\operatorname{\Gamma}\left(\frac{k}{2}+\frac{1}{2}\right)(1-R_{\mathcal{A}}^2)^{\frac{k}{2}-a-\frac{n}{2}+\frac{1}{2}}
\end{equation}
Note that the approximation sign in Strirling's formula above means that the ratio of the two sides converges to one as $x$ goes to infinity.\\

Determination of the lower bound:\\
Observing that $(g+1)^r\geq g^r\geq 0$ for $g\geq 0,r\geq 0$ and further observing that $\operatorname{exp}[g^{-1}(1-R_{\mathcal{A}}^2)^{-1}]\geq 1+g^{-1}(1-R_{\mathcal{A}}^2)^{-1}~\forall g\in\mathbb{R}$ since $\operatorname{exp}(x)\geq 1+x~ \forall x\in\mathbb{R}$, a  lower bound for $p(\mathcal{A}|\mathbf{y},\mathbf{X})$ is given by:
\begin{align}
&p(\mathcal{A}|\mathbf{y},\mathbf{X})/p(\mathcal{A})\\
&\geq \int\limits_{0}^{\infty}g^{-\frac{k}{2}+a+\frac{n}{2}-\frac{3}{2}}\operatorname{exp}\left(-\frac{n}{2g}\right)\left[1+g(1-R_{\mathcal{A}}^2)\right]^{-a-\frac{n}{2}}~dg\\
&=\int\limits_{0}^{\infty}g^{-\frac{k}{2}+a+\frac{n}{2}-\frac{3}{2}}\operatorname{exp}\left(-\frac{n}{2g}\right)\left[g(1-R_{\mathcal{A}}^2)(g^{-1}(1-R_{\mathcal{A}}^2)^{-1}+1)\right]^{-a-\frac{n}{2}}~dg\\
&\geq(1-R_{\mathcal{A}}^2)^{-a-\frac{n}{2}} \int\limits_{0}^{\infty}g^{-\frac{k}{2}-\frac{3}{2}}\operatorname{exp}\left(-\frac{n}{2g}\right)\operatorname{exp}\left(\left(-a-\frac{n}{2}\right)g^{-1}(1-R_{\mathcal{A}}^2)^{-1}\right)~dg\\
&=(1-R_{\mathcal{A}}^2)^{-a-\frac{n}{2}} \int\limits_{0}^{\infty}\left(\frac{1}{g}\right)^{\left(\frac{k}{2}+\frac{1}{2}\right)+1}\operatorname{exp}\left\{-\frac{1}{g}\left[\frac{n}{2}+\left(a+\frac{n}{2}\right)(1-R_{\mathcal{A}}^2)^{-1}\right]\right\}~dg\\
&=(1-R_{\mathcal{A}}^2)^{-a-\frac{n}{2}} \frac{\operatorname{\Gamma}\left(\frac{k}{2}+\frac{1}{2}\right)}{\left[\frac{n}{2}+\left(a+\frac{n}{2}\right)(1-R_{\mathcal{A}}^2)^{-1}\right]^{\frac{k}{2}+\frac{1}{2}}}\\
&=(1-R_{\mathcal{A}}^2)^{-a-\frac{n}{2}}\operatorname{\Gamma}\left(\frac{k}{2}+\frac{1}{2}\right)\left(c_1\frac{n}{2}+c_2\right)^{-\frac{k}{2}-\frac{1}{2}}
\end{align}
Note that (A.21) holds since the function to integrate in (A.20) is proportional to the density of an inverse gamma distribution. Moreover, $c_1$ and $c_2$ denote constants which are independent of $n$.\\

Using the bounds (A.15) and (A.22), an upper bound for the quotient in (A.1) can be obtained for large $n$:
\begin{align}
\frac{p(\mathcal{A}^{\prime}|\mathbf{y},\mathbf{X})}{p(\mathcal{A}|\mathbf{y},\mathbf{X})}&\leq \frac{p(\mathcal{A}^{\prime})\left(\frac{n}{2}\right)^{-\frac{k^{\prime}}{2}-\frac{1}{2}}\operatorname{\Gamma}\left(\frac{k^{\prime}}{2}+\frac{1}{2}\right)(1-R_{\mathcal{A}^{\prime}}^2)^{\frac{k^{\prime}}{2}-a-\frac{n}{2}+\frac{1}{2}}}{p(\mathcal{A})(1-R_{\mathcal{A}}^2)^{-a-\frac{n}{2}}\operatorname{\Gamma}\left(\frac{k}{2}+\frac{1}{2}\right)\left(c_1\frac{n}{2}+c_2\right)^{-\frac{k}{2}-\frac{1}{2}}}\\
&=c_3\frac{\left(c_1\frac{n}{2}+c_2\right)^{\frac{k}{2}+\frac{1}{2}}}{\left(\frac{n}{2}\right)^{\frac{k^{\prime}}{2}+\frac{1}{2}}}\left(\frac{1-R_{\mathcal{A}}^2}{1-R_{\mathcal{A}^{\prime}}^2}\right)^{\frac{n}{2}}\\
&=c_3\frac{\left(c_1\frac{n}{2}+c_2\right)^{\frac{k}{2}+\frac{1}{2}}}{\left(\frac{n}{2}\right)^{\frac{k^{\prime}}{2}+\frac{1}{2}}}\left(\frac{\frac{1}{n}\left[2b+\mathbf{y}^T(\textbf{\text{I}}_n-H_{\mathcal{A}})\mathbf{y}\right]}{\frac{1}{n}\left[2b+\mathbf{y}^T(\textbf{\text{I}}_n-H_{\mathcal{A}^{\prime}})\mathbf{y}\right]}\right)^{\frac{n}{2}}
\end{align}
Note that $c_3$ denotes a constant independent of $n$.

To finally prove model selection consistency, two cases are considered:
\begin{enumerate}[(i)]
\item If $M_{\mathcal{A}}\nsubseteq M_{\mathcal{A}^{\prime}}$ the probability limit of the upper bound (A.25) for $n\rightarrow\infty$ evaluates as follows:
\begin{align}
&\underset{n\rightarrow \infty}{\operatorname{p~lim~}}c_3\frac{\left(c_1\frac{n}{2}+c_2\right)^{\frac{k}{2}+\frac{1}{2}}}{\left(\frac{n}{2}\right)^{\frac{k^{\prime}}{2}+\frac{1}{2}}}\left(\frac{\frac{1}{n}\left[2b+\mathbf{y}^T(\textbf{\text{I}}_n-H_{\mathcal{A}})\mathbf{y}\right]}{\frac{1}{n}\left[2b+\mathbf{y}^T(\textbf{\text{I}}_n-H_{\mathcal{A}^{\prime}})\mathbf{y}\right]}\right)^{\frac{n}{2}}\\
&=\underset{n\rightarrow \infty}{\operatorname{p~lim~}}c_3\frac{\left(c_1\frac{n}{2}+c_2\right)^{\frac{k}{2}+\frac{1}{2}}}{\left(\frac{n}{2}\right)^{\frac{k^{\prime}}{2}+\frac{1}{2}}}\left(\frac{\sigma^2}{\sigma^2+b_{\mathcal{A}^{\prime}}}\right)^{\frac{n}{2}}\\
&=0
\end{align}
Note that equation (A.27) holds by the equations $(15)$ and $(17)$. Further, note that the term $(\sigma^2/(\sigma^2+b_{\mathcal{A}^{\prime}}))^{n/2}$ is element of the interval $(0,1)$ and thus converges exponentially fast to zero. Therefore, the other factor of the product corresponding to equation (A.27) does not influence the limit at all, since it is polynomial in $n$.
\item If $M_{\mathcal{A}}\subseteq M_{\mathcal{A}^{\prime}}$ according to the work of \cite{FERNANDEZ2001} the following holds:
$$
\underset{n\rightarrow \infty}{\operatorname{p~lim~}}\left(\frac{\mathbf{y}^T(\textbf{\text{I}}_n-H_{\mathcal{A}})\mathbf{y}}{\mathbf{y}^T(\textbf{\text{I}}_n-H_{\mathcal{A}^{\prime}})\mathbf{y}}\right)^{\frac{n}{2}}\overset{D}{\longrightarrow}\operatorname{exp}\left(\frac{s}{2}\right)
$$
where $s$ has a $\mbox{\large$\chi$}^2$ distribution with $k^{\prime}-k$ degrees of freedom and $\overset{D}{\longrightarrow}$ denotes convergence in distribution. Thus, for $n\rightarrow\infty$ the limit of the upper bound (A.25) evaluates as follows:
\begin{align}
&\underset{n\rightarrow \infty}{\operatorname{p~lim~}}c_3\frac{\left(c_1\frac{n}{2}+c_2\right)^{\frac{k}{2}+\frac{1}{2}}}{\left(\frac{n}{2}\right)^{\frac{k^{\prime}}{2}+\frac{1}{2}}}\left(\frac{\frac{1}{n}\left[2b+\mathbf{y}^T(\textbf{\text{I}}_n-H_{\mathcal{A}})\mathbf{y}\right]}{\frac{1}{n}\left[2b+\mathbf{y}^T(\textbf{\text{I}}_n-H_{\mathcal{A}^{\prime}})\mathbf{y}\right]}\right)^{\frac{n}{2}}\\
&=\underset{n\rightarrow \infty}{\operatorname{p~lim~}}c_3\frac{\left(c_1\frac{n}{2}+c_2\right)^{\frac{k}{2}+\frac{1}{2}}}{\left(\frac{n}{2}\right)^{\frac{k^{\prime}}{2}+\frac{1}{2}}}\operatorname{exp}\left(\frac{s}{2}\right)\\
&=0
\end{align}
Note that $k^{\prime}>k$ since $M_{\mathcal{A}}\neq M_{\mathcal{A}^{\prime}}$.
\end{enumerate}

Finally, in both cases the upper bound of the quotient in (A.1) converges to $0$ and thus also the quotient itself since it is greater or equal to $0$ and the proof is completed. 

\hfill$\square$


\bibliography{literature}

\end{document}